\documentclass[lettersize,journal]{IEEEtran}
\IEEEoverridecommandlockouts
\usepackage{times}
\usepackage{amsmath,dsfont}
\usepackage{amssymb,amsthm}
\usepackage{epsfig,verbatim}
\usepackage{subcaption}
\usepackage{setspace}
\usepackage{color}
\usepackage{bm}
\usepackage{cite}
\usepackage{epstopdf}
\usepackage{graphics}
\usepackage{accents}
\usepackage{acronym}
\usepackage{booktabs}
\usepackage{mathtools} 
\usepackage{enumitem}
\usepackage{multirow}
\usepackage{dblfloatfix}
\usepackage{amsmath}
\usepackage{gensymb}
\usepackage[flushleft]{threeparttable}
\usepackage{algpseudocode}
\usepackage{bbm}
\usepackage[bookmarks,colorlinks]{hyperref}

\makeatletter
\newcommand{\removelatexerror}{\let\@latex@error\@gobble}
\makeatother

\usepackage[ruled,linesnumbered,noend]{algorithm2e}  %,algochapter
\newcommand{\myVec}[1]{{\boldsymbol{#1}}}
\newcommand{\myMat}[1]{{\boldsymbol{#1}}}
\newcommand{\mySet}[1]{\mathcal{#1}}
\newcommand*\rfrac[2]{{}^{#1}\!/_{#2}}%running fraction with slash - requires math mode.
\newcommand{\Weights}{\myVec{\varphi}}

\newcommand{\Blklen}{T}		
\newcommand{\Bitslen}{C}% observations length
\newcommand{\Codlen}{C}			 			% observations length

\newcommand{\Vecdim}[1]{} %\newcommand{\Vecdim}[1]{^{#1}}				% vector dimensions

\DeclareMathOperator\arctanh{arctanh}

\definecolor{NewColor}{rgb}{0,0,0} %{0.2,0,0.5}

\acrodef{fc}[FC]{fully connected}
\acrodef{adc}[ADC]{analog-to-digital convertor}
\acrodef{cs}[CS]{compressed sensing}
\acrodef{dtft}[DTFT]{discrete-time Fourier transform}
\acrodef{dnn}[DNN]{deep neural network} 
\acrodef{csi}[CSI]{channel state information}
\acrodef{bpsk}[BPSK]{binary phase shift keying}
\acrodef{map}[MAP]{maximum a-posteriori probability}
\acrodef{snr}[SNR]{signal-noise ratio}
\acrodef{bs}[BS]{base station} 
\acrodef{iot}[IOT]{Interent of Things}
\acrodef{mimo}[MIMO]{multiple-input multiple-output}
\acrodef{siso}[SISO]{single-input single-output}
\acrodef{mse}[MSE]{mean-squared error}
\acrodef{pdf}[PDF]{probability density function}
\acrodef{rv}[RV]{random variable}
\acrodef{ml}[ML]{machine learning}
\acrodef{mc}[MC]{monte carlo}
\acrodef{fec}[FEC]{forward error correction}
\acrodef{qpsk}[QPSK]{quadrature phase shift keying}
\acrodef{rs}[RS]{Reed-Solomon}
\acrodef{lti}[LTI]{linear time-invariant}
\acrodef{wss}[WSS]{wide-sense stationary}
\acrodef{psd}[PSD]{power spectral density}
\acrodef{ser}[SER]{symbol error rate} 
\acrodef{ber}[BER]{bit error rate} 
\acrodef{gd}[GD]{gradient descent}
\acrodef{sgd}[SGD]{stochastic gradient descent} 
\acrodef{isi}[ISI]{intersymbol interference}  
\acrodef{awgn}[AWGN]{additive zero-mean white complex Gaussian noise} 
\acrodef{ut}[UT]{user terminal} 
\acrodef{mmw}[mmWave]{millimeter wave}
\acrodef{noma}[NOMA]{non-orthogonal multiple access}
\acrodef{mac}[MAC]{mulitple access channel}
\acrodef{fl}[FL]{Federated learning}
\acrodef{lstm}[LSTM]{long short-term memory}
\acrodef{maml}[MAML]{model-agnostic meta-learning}
\acrodef{sic}[SIC]{soft interference cancellation}
\acrodef{pmf}[PMF]{probability mass function}
\acrodef{crc}[CRC]{cyclic redundancy check}
\acrodef{ece}[ECE]{expected calibration error}
\acrodef{lbd}[LBD]{learnable Bernoulli dropout}
\acrodef{arm}[ARM]{Augment-REINFORCE-Merge}
\acrodef{kl}[KL]{Kullback-Leibler}
\acrodef{ai}[AI]{artificial intelligence}
\acrodef{bp}[BP]{belief propagation}
\acrodef{wbp}[WBP]{weighted belief propagation}
\acrodef{llr}[LLR]{log likelihood ratio}
\acrodef{ecc}[ECC]{error-correction code}

\IEEEoverridecommandlockouts 

\begin{document}

\title{Uncertainty-Aware and Reliable Neural MIMO Receivers via Modular Bayesian Deep Learning}
\author{  
	\IEEEauthorblockN{Tomer Raviv, Sangwoo Park, Osvaldo Simeone, and Nir Shlezinger\\
	} 
	\thanks{
 Parts of this work were presented at the 2023 IEEE International Conference on Communications (ICC) as the paper~\cite{raviv2023modular}.
	   T. Raviv and N. Shlezinger are with the School of ECE, Ben-Gurion University of the Negev, Beer-Sheva, Israel (e-mail: tomerraviv95@gmail.com, nirshl@bgu.ac.il).
		S. Park and O. Simeone are with King's Communication Learning \& Information Processing (KCLIP) lab, Centre for Intelligent Information Processing Systems (CIIPS),  at the Department of Engineering, King’s College London,  U.K. (email: \{sangwoo.park; osvaldo.simeone\}@kcl.ac.uk). 
  	The work of T. Raviv and N. Shlezinger was partially supported by the Israeli Innovation Authority. 
  The work of O. Simeone was partially supported by the European Union’s Horizon Europe project CENTRIC (101096379),  by an Open Fellowship of the EPSRC (EP/W024101/1),  by the EPSRC project (EP/X011852/1), and by  Project REASON, a UK Government funded project under the Future Open Networks Research Challenge (FONRC) sponsored by the Department of Science Innovation and Technology (DSIT).}
	\vspace{-0.5cm}
}

\maketitle
	
	\maketitle

	\pagestyle{plain}
	\thispagestyle{plain}
	%----------------------------------------------------------------------------------------
	%	ABSTRACT
	%----------------------------------------------------------------------------------------
	\begin{abstract} 
Deep learning is envisioned to play a key role in the design of future wireless receivers. 
A popular approach to design learning-aided receivers combines  \acp{dnn} with traditional model-based receiver algorithms, realizing  hybrid model-based data-driven architectures. % can harness the abstractness and learning capabiliteis of \acp{dnn} while accounting for domain knowledge and established processing steps.
Such architectures typically include multiple modules, each carrying out a different functionality dictated by the model-based receiver workflow. Conventionally trained \ac{dnn}-based modules are known to produce poorly calibrated, typically overconfident, decisions. Consequently, incorrect decisions may propagate through the architecture without any indication of their insufficient accuracy. To address this problem, we present a novel combination of Bayesian deep learning with hybrid model-based data-driven architectures for wireless  receiver design. The proposed methodology, referred to as {\em modular Bayesian deep learning}, is designed to yield calibrated modules, which in turn improves both accuracy and calibration of the overall receiver. We specialize this approach for two fundamental tasks in \ac{mimo} receivers -- equalization and decoding. In the presence of scarce data, the ability of  modular Bayesian deep learning to produce reliable uncertainty measures is consistently shown to directly translate into improved performance of the overall \ac{mimo} receiver chain.
\end{abstract}
 \acresetall
	%----------------------------------------------------------------------------------------
	%	Introduction
	%----------------------------------------------------------------------------------------
	\section{Introduction} \label{sec:intro}
%\vspace{-0.1cm} 

Wireless communication technologies are subject to increasing demands for connectivity and throughput. These are driven by the growth in media transmissions like  videos, augmented reality and virtual reality, as well as in applications such as autonomous driving and smart manufacturing~\cite{wang2023road}.   To support the growing demands and new applications, emerging technologies like  THz communication~\cite{sarieddeen2021overview}, holographic \ac{mimo}~\cite{gong2023holographic}, intelligent surfaces~\cite{basar2019wireless}, spectrum sharing~\cite{chepuri2023integrated},  and novel antennas~\cite{shlezinger2021dynamic} are being explored, though adding significant design and operational complexities~\cite{raviv2023adaptive}. %For example, holographic MIMO hardware can introduce non-linearities, IRS complicates channel estimation, and classical communication models may not be applicable in new settings like mmWave and THz spectrums due to far-field assumption violations and lossy propagation.}

\Acp{dnn} were shown to successfully learn complex tasks from data in domains such as computer vision and natural language processing. This dramatic success has led to a burgeoning interest in the design of \ac{dnn}-aided  communications receivers \cite{gunduz2019machine,simeone2018very,balatsoukas2019deep, farsad2020data,mao2018deep}. \ac{dnn}-aided receivers can  operate efficiently in scenarios where the channel model is unknown, highly complex, or difficult to optimize \cite{dai2020deep}. The incorporation of deep learning into receiver processing is thus often envisioned to be  an important contributor to meeting the rapidly increasing demands of wireless systems  in  throughput, coverage, and robustness~\cite{saad2019vision}.

Despite their promise in facilitating physical layer communications,  particularly or \ac{mimo} systems, the deployment of \acp{dnn} faces several challenges that may limit their direct applicability to wireless receivers. The dynamic nature of wireless channels combined with the limited computational and memory resources of wireless devices indicates that \ac{mimo}  receiver processing is fundamentally distinct from conventional deep learning domains, such as computer vision and natural language processing~\cite{raviv2023adaptive,chen2023learning}. Specifically, the necessity for online training may arise due to the time-varying settings of wireless communications. Nonetheless, pilot data corresponding to new channels is usually scarce. \acp{dnn} trained using conventional deep learning mechanisms, i.e., via {\em frequentist learning}, are prone to overfitting in such settings~\cite{jose2020free}. Moreover, they are likely to  yield poorly calibrated outputs~\cite{guo2017calibration,simeone2022machine}, e.g., produce wrong decisions with high confidence. 

%This implies that a \ac{dnn}  trained for a particular channel may not perform well on future channel realizations, requiring frequent transmissions of training (pilot) data for retraining to achieve satisfactory performance. Furthermore, as typical \acp{dnn} are highly-parameterized in order to represent a broad range of mappings, they require a large amount of training data to avoid \emph{overfitting} \cite{jose2020free}. \acp{dnn} that suffer from overfitting generally produce wrong decisions with high confidence at deployment \cite{simeone2022machine}. Overconfident decisions may lead to poor trustworthiness properties as well as poor performance of downstream tasks that rely on these values, such as soft decoding.

A candidate approach to design   \ac{dnn}-aided receivers that can be trained with limited data incorporates model-based receiver algorithms as a form of inductive bias into a \emph{hybrid model-based data-driven architecture} \cite{shlezinger2020model, shlezinger2022model, shlezinger2023model}. One such model-based deep learning methodology is {\em deep unfolding}~\cite{monga2021algorithm}. Deep unfolding converts an iterative optimizer into a discriminative machine learning system by introducing trainable parameters within each of a fixed number of iterations \cite{shlezinger2023model,shlezinger2023discriminative}. Deep unfolded architectures treat each iteration as a trainable module, obtaining a  modular and interpretable \ac{dnn} architecture \cite{raviv2022online}, with each module carrying out a specific functionality within a conventional receiver processing algorithm. Modularity is informed by  domain knowledge, potentially enhancing the generalization capability of \ac{dnn}-based receivers  \cite{shlezinger2020data,shlezinger2019deepSIC, shlezinger2019viterbinet,raviv2020data,van2022deep,jiang2021ai}.

% An alternative complementary approach aims at generating and enriching data available to \ac{dnn}-aided receivers, via self-supervision~\cite{raviv2022online} or data augmentation~\cite{raviv2022data}. These existing approaches, which focus on the {\em architecture} or the {\em data}, often alleviate the effect of limited data on the receiver accuracy, i.e., they allow the transmitted symbols to be detected with high accuracy. Yet, they still share the limitations of \acp{dnn} in producing overconfident decisions~\cite{ovadia2019can}. Overconfident decisions, the most common form of {\em miscalibration}, may lead to poor trustworthiness properties as well as poor performance of downstream tasks that rely on these values, such as soft decoding. 

%While the hybrid model-based/data-driven receivers utilize data-driven deep learning to alleviate the  practical limitations of existing communication models, e.g., perfect channel knowledge, the data-driven part of the hybrid receiver may still suffer from overconfident decisions unless certain amount of training data is provided.

%The need to ensure \emph{reliable}, i.e., \emph{well-calibrated}, predictions is typically treated in the machine learning research via the {\em learning} procedure \cite{guo2017calibration, blundell2015weight,gal2016dropout, boluki2020learnable}. 

Hybrid model-based data-driven methods require less data to reliably train compared to black box architectures. Yet, the resulting system may be negatively affected by the inclusion of conventionally trained \ac{dnn}-based modules, i.e., \acp{dnn} trained via  frequentist learning~\cite{simeone2022machine}. This follows from the frequent overconfident decisions that are often produced by  such  conventional \acp{dnn}, particularly when learning from limited data sets.
Such poorly calibrated  \ac{dnn}-based modules may propagate inaccurate predictions through the architecture without any indication of their insufficient accuracy. 

\begin{figure*}[t]
	\centering
	\includegraphics[width = 0.72\textwidth]{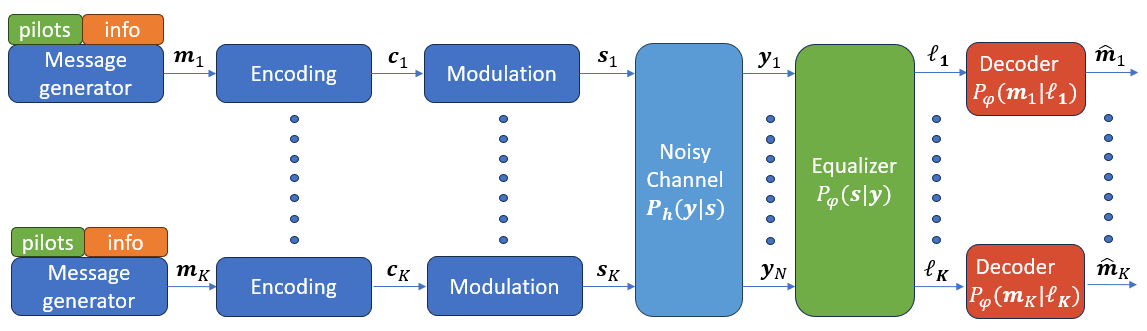}
	\caption{The communication system studied in this paper.} 
 % \textcolor{red}{there is something wrong with this figure. Why are there $K$ channel outputs? I am also missing the vector $\myVec{h}$ and the distribution representing the noisy channel. Also the vectors $\hat{P}$ should have the user index as their subscript, and the decoder should be done for each user separately. The time index $i$ should be added where relevant...}}
	\label{fig:communication_system}
\end{figure*}

The machine learning framework of \emph{Bayesian learning} accommodates learning techniques with improved calibration compared with frequentist learning~\cite{jospin2022hands,chang2021bayesian,wang2020survey,simeone2022machine}. In Bayesian deep learning, the parameters  of a \ac{dnn} are treated as random variables. Doing so provides means to quantify the \emph{epistemic} uncertainty, which arises due to lack of training data. Bayesian deep learning was shown to improve the reliability of \acp{dnn} in wireless communication systems \cite{zecchin2023robust, cohen2022bayesian}. Nonetheless, these existing studies adopt \emph{black-box} \ac{dnn} architectures. Thus, the existing approaches do not leverage the modularity of model-based data-driven receivers, and do not account for the ability to assign interpretable meaning to intermediate features exchanged in model-based deep learning architectures. 

In this work we present a novel methodology, referred to as \emph{modular Bayesian deep learning}, that  combines Bayesian learning with hybrid model-based data-driven architectures for \ac{dnn}-aided receivers.  The proposed approach calibrates the internal DNN-based modules,  rather than merely calibrating the overall system output. This has the advantage of yielding well-calibrated soft outputs that can  benefit downstream tasks, as we  empirically demonstrate.

Our main contributions are summarized as follows.
\begin{itemize}
    \item \textbf{Modular Bayesian deep learning:} We introduce a modular  Bayesian deep learning framework that integrates Bayesian learning with model-based deep learning. Accounting for the modular architecture of hybrid model-based data-driven receivers, Bayesian learning is applied on a per-module basis, enhancing the calibration of soft estimates exchanged between internal \ac{dnn} modules. The training procedure is based on dedicated {\em sequential Bayesian learning} operation, which  boosts calibration of internal modules while encouraging subsequent modules to exploit these calibrated features.  %The proposed scheme enhances the performance of the receiver in both terms of hard detection (accuracy) and soft decision (calibration) and is formulated for the DeepSIC architecture. %the calibration offered by Bayesian learning by addressing each sub-module in the model itself, and is formulated for the DeepSIC architecture.
    \item \textbf{Modular Bayesian deep \ac{mimo} receivers:} 
We specialize the proposed methodology to \ac{mimo} receivers. To this end, we adopt two established \ac{dnn}-aided architectures -- the DeepSIC \ac{mimo} equalizer of \cite{shlezinger2019deepSIC} and the deep unfolded \ac{wbp} soft decoder of \cite{nachmani2018deep} -- and derive their design as modular Bayesian deep learning systems.  
    \item \textbf{Extensive experimentation:} We evaluate the advantages of modular Bayesian learning for  deep \ac{mimo}  receivers based on DeepSIC and \ac{wbp}. Our evaluation first considers equalization and decoding separately, followed by an evaluation of the overall receiver chain composed of the cascade of equalization and decoding.  The proposed Bayesian modular approach is systematically shown to outperform conventional  frequentist learning, as well as traditional Bayesian black-box learning, while being reliable and interpretable.
\end{itemize}

The remainder of this paper is organized as follows: Section~\ref{sec:background} formulates the considered \ac{mimo} system model, and reviews some necessary preliminaries on Bayesian learning and model-based deep receivers. Section~\ref{sec:model_based_bayesian_learning} presents the proposed methodology of modular Bayesian learning, and specializes it for \ac{mimo} equalization using DeepSIC and for decoding using the \ac{wbp}. Our extensive numerical experiments are reported in Section~\ref{sec:Simulation}, while Section~\ref{sec:conclusion} provides concluding remarks. 

Throughout the paper, we use boldface letters for multivariate quantities, e.g., ${\myVec{x}}$ is a vector; 
%upper-cased boldface letters for matrices, e.g., $\myMat{X}$; 
calligraphic letters, such as $\mySet{X}$ for sets, with $|\mySet{X}|$ being the cardinality of $\mySet{X}$; and  $\mySet{R}$ for the set of real numbers. The notation $({\cdot})^{T}$ stands for the transpose operation.
	%----------------------------------------------------------------------------------------
	%	System Model
	%----------------------------------------------------------------------------------------
	\section{System Model and Preliminaries}
\label{sec:background}

In this section, we describe the \ac{mimo} communication system under study in Subsection~\ref{subsec:system_model}, and  review the conventional frequentist learning of \ac{mimo} receivers in Subsection~\ref{subsec:frequentist}. Then, we briefly present basics of model-based learning for deep receivers in Subsection~\ref{subsec:architecture}, and review our main running examples of DeepSIC~\cite{shlezinger2019deepSIC} and \ac{wbp}~\cite{nachmani2018deep}. 

\subsection{System Model}
\label{subsec:system_model}

We consider an uplink \ac{mimo} digital communication system with $K$ single-antenna users. The users transmit encoded messages, independently of each other, to a base station equipped with $N$ antennas.
%
% \smallskip
% {\bf Transmission Chain:} 
Each user of index $k$ uses a linear block error-correction code to encode a length $m$ binary message  ${\bm{m}_k\in\{0,1\}^{m}}$. Encoding is carried out by multiplication  with a generator matrix $\bm{G}$, resulting in a length $\Codlen \geq m$ codeword given by ${\bm{c}_k = \bm{G}^{T}\bm{m}_k \in\{0,1\}^{\Codlen}}$. The same codebook is employed for all users. 
The messages are modulated, such that each symbol ${s}_k[i]$ at the $i$th time index is a member of the set of constellation points $\mySet{S}$. 
The  transmitted symbols vector form the  $K\times1$ vector $\myVec{s}[i] = [s_1[i]  \ldots s_K[i]] \in \mySet{S}^K$. 

% \smallskip
% {\bf Channel Model:} 
We adopt a block-fading model, i.e., the channel  remains constant within a block of duration dictated by the coherence  time of the channel. A single block of duration $\Blklen^{\rm tran}$ is generally divided into two parts: the pilot part, composed of $\Blklen^{\rm pilot}$ pilots, and the message part of $\Blklen^{\rm info} = \Blklen^{\rm tran}-\Blklen^{\rm pilot}$ information symbols. The corresponding indices for transmitted pilot symbols and information symbols are written as $\mySet{\Blklen}^{\rm pilot} = \{1,...,\Blklen^{\rm pilot}\}$, and $\mySet{\Blklen}^{\rm info} = \{\Blklen^{\rm pilot}+1,...,\Blklen^{\rm tran}\}$, respectively.

To accommodate complex and non-linear channel models, we represent the channel input-output relationship by a generic conditional distribution. Accordingly, the corresponding received signal vector $\myVec{y}[i] \in \mySet{R}^N$ is determined as
\begin{align}
\label{eq:general_channel_mapping}
    \myVec{y}[i] \sim P_{\myVec{h}}(\myVec{y}[i]|\myVec{s}[i]),
\end{align}
which is subject to the unknown conditional  distribution $P_{\myVec{h}}(\cdot|\cdot)$  that depends on the current channel parameters $\myVec{h}$. 

The received channel outputs in \eqref{eq:general_channel_mapping} are processed by the receiver, whose goal is to recover correctly each message. This estimate of the $k$th user message is denoted by $\hat{\myVec{m}}_k$. 
This mapping can be generally divided into two different tasks: $(\emph{i})$ equalization and $(\emph{ii})$ decoding. Equalization maps the received signal $\boldsymbol{y}[i]$ as an input into a vector of soft symbols estimates for each user, denoted $\hat{\boldsymbol{P}}_1[i],\ldots,\hat{\boldsymbol{P}}_K[i]$. Decoding is done for each user separately, obtaining an estimation $\hat{\myVec{m}}_k$ of the transmitted message from the soft probability vectors $\{\hat{\boldsymbol{P}}_k[i]\}_i$ corresponding to the codeword. The latter often involves first obtaining a soft estimate for each bit in the message, where we use $\hat{L}_k[v]$ to denote the estimated probability that the $v$th bit in the $k$th message is one. 
The entire communication chain is illustrated in Fig.~\ref{fig:communication_system}.

\subsection{Frequentist Deep Receivers}
\label{subsec:frequentist}

Deep receivers employ \acp{dnn} to map the block of channel output into the transmitted message, i.e., as part of the equalization and/or decoding operations. 
In conventional {\em frequentist} learning, optimization of the parameter vector $\Weights$ dictating the receiver mappings is typically done by minimizing the cross-entropy loss using a labelled pair of inputs and outputs  \cite{simeone2018very}, i.e., 
\begin{align}
    \Weights^{\rm freq} = \arg \min_{\Weights} \mySet{L}_{\rm CE}(\Weights).
    \label{eq:frequentist_general_sol}
\end{align}

For example, to optimize an equalizer (formulated here for a single user case for brevity), one employs $\mySet{\Blklen}^{\rm pilot}$ pilot symbols and their corresponding observed channel outputs to compute 
\begin{align}
    \label{eq:ce_loss}
    \mySet{L}_{\rm CE}(\Weights) = - \frac{1}{|\mySet{\Blklen}^{\rm pilot}|}\sum_{i \in \mySet{\Blklen}^{\rm pilot}}\log P_{\Weights}(\boldsymbol{s}[i]|\boldsymbol{y}[i]).
\end{align}
In \eqref{eq:ce_loss}, $P_{\Weights}(\boldsymbol{s}[i]|\boldsymbol{y}[i])$ is the soft estimate associated with the symbol $\boldsymbol{s}[i]$ computed from $\boldsymbol{y}[i]$ with a \ac{dnn}-aided equalizer with parameters $\Weights$. When such an equalizer is used, the stacking of $\{P_{\Weights}(\boldsymbol{s}|\boldsymbol{y}[i])\}_{\boldsymbol{s} \in \mathcal{S}}$ forms the   soft estimate $\hat{\boldsymbol{P}}[i]$. 

Calibration is based on the reliability of its soft estimates. In particular, a prediction model $P_{\Weights}(\myVec{s}[i]|\myVec{y}[i])$ as in the above \ac{dnn}-aided equalizer example is \emph{perfectly calibrated} if the conditional distribution of the transmitted symbols given the predictor's output satisfies the equality \cite{guo2017calibration}  
\begin{align}
    \label{eq:perfect_cal}
    P(\myVec{s}[i]=\hat{\myVec{s}}[i] | P_{\Weights}(\hat{\myVec{s}}[i]|\myVec{y}[i]) = \pi )=\pi \text{ for all } \pi \in [0,1].
\end{align}
When \eqref{eq:perfect_cal} holds, then the soft outputs of the prediction model coincide with the true underlying conditional distribution. 

% After optimization, the trained model is used to classify data samples $i \in \mySet{\Blklen}^{\rm info}$ by
% \begin{align}
% \label{eq:freq_receiver}
% \hat{\boldsymbol{s}}[i] = \arg \max_ {\boldsymbol{s}[i] \in \mySet{S}^{K}} P_{\Weights^{\rm freq}}(\boldsymbol{s}[i]|\boldsymbol{y}[i]).
% \end{align}

% In the presence of limited data, the conventional deep receiver in \eqref{eq:freq_receiver} tends to suffer from overfitting when the number of parameters (the cardinality of $\Weights$) is large. This may cause a degraded accuracy, i.e., a high probability that $\hat{x}^\text{out} \neq x^\text{out}$, at test time. Furthermore, the trained model \eqref{eq:freq_receiver} tends to produce overconfident predictions. This is in the sense that the confidence level $P_{\Weights^{\rm freq}}(x^\text{out}|x^\text{in})$ produced for the prediction $x^\text{out}$ is typically larger than the corresponding actual test accuracy. Overfitting can be alleviated via the adoption of a hybrid model-based/data-driven architecture, while poor calibration can be addressed via Bayesian learning (see Section~\ref{sec:model_based_bayesian_learning}). We next elaborate on these individual approaches, starting with the former.

\begin{figure}
\removelatexerror
  \begin{algorithm}[H]
    \caption{DeepSIC Inference}
    \label{alg:deepsic}
    \SetAlgoLined
    \SetKwInOut{Input}{Input}
        \Input{Channel output $\boldsymbol{y}[i]$; Parameter vector $\Weights^{\rm SIC}$.} %  \newline 
    \SetKwInOut{Output}{Output}
    \SetKwProg{DeepSICInference}{DeepSIC Inference}{}{}
    \Output{Soft estimations for all users $\{\hat{\myVec{P}}_k[i]\}_{k=1}^K$.}    \DeepSICInference{}{%$(\boldsymbol{y},\Weights^{\rm SIC})$}{
    
    Generate an initial guess $\{\hat{\myVec{P}}^{(k,0)}\}_{k=1}^K$.
    
    \For{$q\in\{1,...,Q\}$\label{line:loop}}{ 
            \For{$k\in\{1,...,K\}$}{ 
                 Calculate $\hat{\myVec{P}}^{(k,q)}$ by \eqref{eq:soft_freq_out}.
            }            
        }
    \KwRet{$\{\hat{\myVec{P}}_k[i]=\hat{\myVec{P}}^{(k,Q)}\}_{k=1}^K$} \label{line:return}
  }
  \end{algorithm}
\end{figure}

\subsection{Hybrid Model-Based Data-Driven Architectures}
\label{subsec:architecture}

As mentioned in Section \ref{sec:intro}, two representative examples of hybrid model-based data-driven architectures employed by deep receivers are: $(\emph{i})$ the DeepSIC equalizer of \cite{shlezinger2019deepSIC}; and $(\emph{ii})$ the unfolded \ac{wbp} introduced in \cite{nachmani2016learning,nachmani2018deep} for channel decoding. As these architectures are employed as our main running examples in the sequel, we next briefly recall their formuation.

\subsubsection{DeepSIC}
\label{subsubsec:deepsic}

\begin{figure}
	\centering
	\includegraphics[width = 0.9\columnwidth]{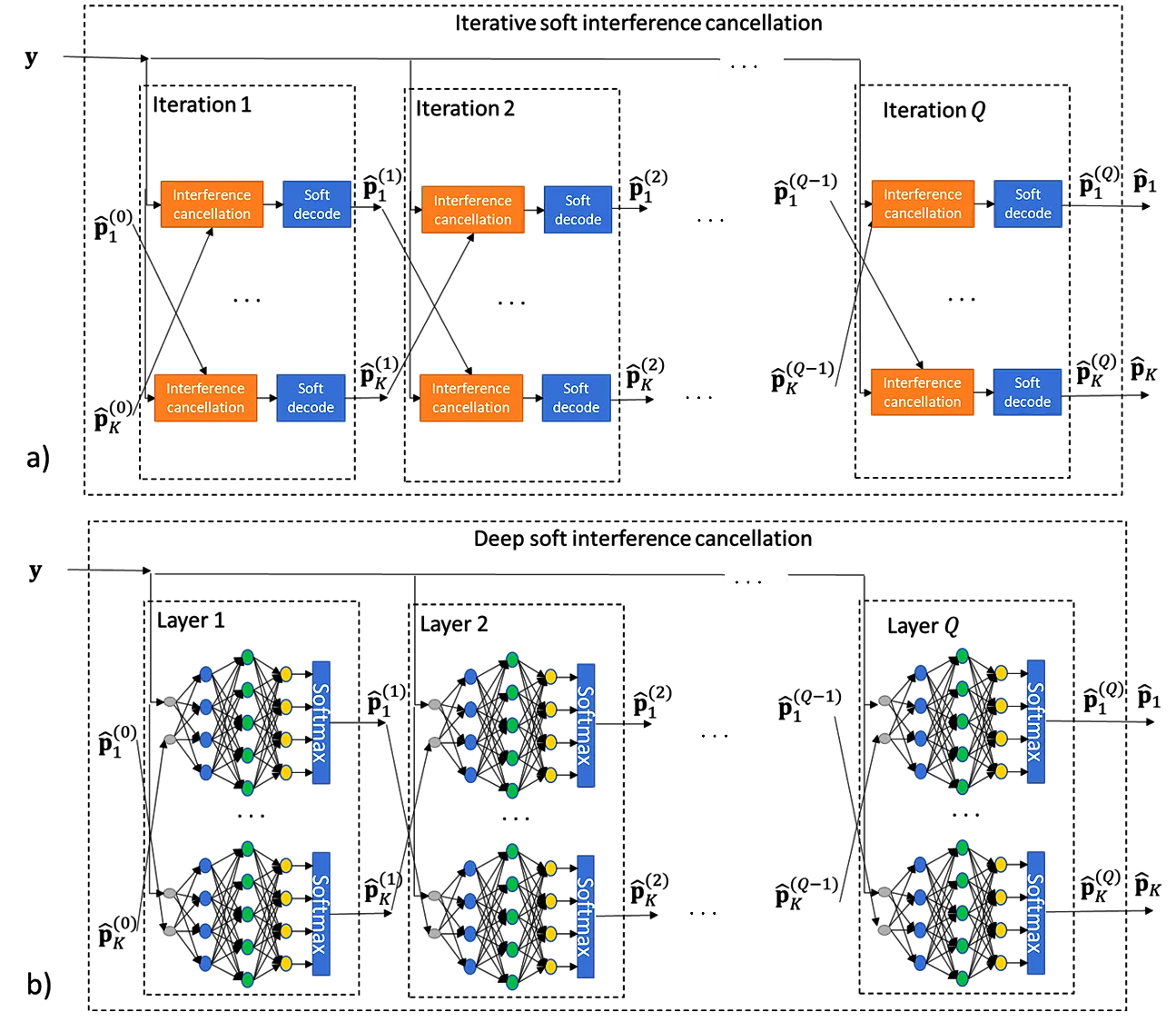}
	\caption{$(a)$ Iterative \acs{sic}; and $(b)$ DeepSIC \cite{shlezinger2019deepSIC}. In this paper, we introduce modular Bayesian learning, which calibrates the internal modules of the DeepSIC architecture with the double goal of improving end-to-end accuracy and of producing  well-calibrated soft outputs for the  benefit of downstream tasks.}
	%	\vspace{-0.4cm}
	\label{fig:SoftIC}
\end{figure}

The  DeepSIC architecture proposed in \cite{shlezinger2019deepSIC} is a modular model-based \ac{dnn}-aided \ac{mimo} demodulator, based on the classic \ac{sic} algorithm~\cite{choi2000iterative}. \ac{sic} operates in $Q$ iterations. This equalization algorithm refines an estimate of the conditional \acl{pmf} for the constellation points in $\mySet{S}$ denoted by $\hat{\myVec{P}}^{(k,q)}$, for each symbol transmitted by a user of index $k$,  throughout the iterations $q=1,\
\ldots,Q$. The estimate is generated for every user $k \in \{1,\ldots, K\}$ and every iteration $q \in \{1,\ldots, Q\}$ by using the soft estimates $\{\hat{\myVec{P}}^{(l,q-1)}\}_{l\neq k}$ of the interfering symbols $\{\boldsymbol{s}_l [i]\}_{l \neq k}$ obtained in the previous iteration $q-1$ as well as the channel output $\myVec{y}[i]$. This procedure is illustrated in Fig.~\ref{fig:SoftIC}(a).

In DeepSIC, interference cancellation and soft detection steps are implemented with \acp{dnn}. Specifically, the soft estimate of the symbol transmitted by $k$th user in the $q$th iteration is calculated by a \ac{dnn}-based classifying module with parameter vector $\Weights^{(k,q)}$. Accordingly, the $|\mySet{S}| \times 1$ probability mass function vector for user $k$ at iteration $q$ under time index $i$ can be obtained as
\begin{align}
    \label{eq:soft_freq_out}
    \hat{\myVec{P}}^{(k,q)} = \Big\{P_{\Weights^{(k,q)}}(\myVec{s}|\myVec{y}[i], \{ \hat{\myVec{P}}^{(l,q-1)}[i] \}_{l\neq k}  )\Big\}_{\myVec{s}\in\mathcal{S}},
\end{align} 
where DeepSIC is parameterized by the set of model parameter vectors  $\Weights^{\rm SIC} = \{\{ \Weights^{(k,q)} \}_{k=1}^K\}_{q=1}^Q$. The entire DeepSIC architecture is illustrated in Fig.~\ref{fig:SoftIC}(b) and its inference procedure is summarized in Algorithm~\ref{alg:deepsic}.

In the original formulation in \cite{shlezinger2019deepSIC}, DeepSIC is trained in a frequentist manner, i.e., by minimizing the cross-entropy loss \eqref{eq:ce_loss}. By treating each user $k=1,\ldots,K$ equally, the resulting loss is given by
\begin{align}
    \label{eq:deepsic_loss}
    \mySet{L}_{\rm CE}^{\rm SIC}(\Weights) = \sum_{k=1}^K \mySet{L}_{\rm CE}^{\rm MSIC}(\Weights^{(k,Q)}),
\end{align}
where the user-wise cross-entropy is evaluated over the pilot set  $ \mySet{\Blklen}^{\rm pilot}$ as
\begin{align}
   & \mySet{L}_{\rm CE}^{\rm MSIC}(\Weights^{(k,q)}) = - \frac{1}{|\mySet{\Blklen}^{\rm pilot}|}\notag \\
   &\quad\times \sum_{i\in  \mySet{\Blklen}^{\rm pilot}} \log P_{\Weights^{(k,q)}}(\myVec{s}_k[i]|\myVec{y}[i], \{ \hat{\myVec{P}}^{(l,q-1)}[i] \}_{l\neq k}).
    \label{eq:module_deepsic_loss}
\end{align}

In \cite{shlezinger2019deepSIC}, the soft estimates for each user of index $k$ were obtained via the final soft output, i.e., $\hat{\boldsymbol{P}}_k[i]$ is obtained as $\hat{\myVec{P}}^{(k,Q)}$. 
It was shown in \cite{shlezinger2019deepSIC} that under such setting,
DeepSIC can outperform frequentist black-box deep receivers under limited pilots scenarios. However, its calibration performance is not discussed.

\subsubsection{Weighted Belief Propagation}
\label{subsubsec:wbp}

\Ac{bp} \cite{pearl2014probabilistic} is an efficient algorithm for  calculating the marginal probabilities of a set of random variables whose relationship can be represented as a graph. A graphical representation used in channel coding is the Tanner graph  of  linear error-correcting codes \cite{tomlinson2017error}. A Tanner graph is a bipartite graph with two types of nodes: variable nodes and check nodes. The variable nodes correspond to  bits of the encoded message, and the check nodes represent  constraints 
 that the  message must satisfy.

In the presence of loops in the graph,  \ac{bp}  is generally suboptimal. Its performance can be improved by weighting the messages via a learnable variant of the standard \ac{bp}, called \ac{wbp} \cite{nachmani2016learning, nachmani2018deep}. In \ac{wbp}, the messages passed between nodes in the graphical model are weighted based on the relative soft-information they represent. These weights are learned from data by unfolding the message passing iterations of \ac{bp} into a machine learning architecture. In particular, the message passing algorithm operates by iteratively passing messages over edges from variables nodes to check nodes and vice versa.

As the decoder is applied to each user seperately, we omit the user index $k$ in the following.
To formulate the decoder, let the WBP message from variable node $v$ to check node $h$ at iteration $q$ be denoted by $M^{(q)}_{v \rightarrow h}$, and from check node $h$ to variable node $v$ by $M^{(q)}_{h \rightarrow v}$. The messages are initialized $M^{(q)}_{v \rightarrow h} = M^{(q)}_{h \rightarrow v} = 0$ for all $(v,h) \in \mathcal{H}$, where $\mathcal{H}$ denotes all valid variable-check connections. For instance, validity is determined for linear block codes by the element in the $v$th row and $h$th column in the parity check matrix. The input \ac{llr} values to the decoder are initialized from the outputs of the detection phase. 

To this end, one first finds the soft estimate $\tilde{\boldsymbol{c}}\in\mySet{R}^C$ of the codeword $\boldsymbol{c}$. Given the $i$th soft symbol probability $\hat{\boldsymbol{P}}[i]$, the soft estimate $\tilde{\boldsymbol{c}}$ can be found by translating $\hat{\boldsymbol{P}}[i]$ to $r = \log_2|\mathcal{S}|$ soft bits probabilities $\{\tilde{\boldsymbol{c}}[i\cdot r],...,\tilde{\boldsymbol{c}}[i\cdot r+r-1]\}$ with $\tilde{\boldsymbol{c}}[v] = P(\boldsymbol{c}[v] = 1| \hat{\boldsymbol{P}}[i])$ for $v = i \cdot r,...,i\cdot r+r-1$. Then the \acp{llr} are computed by
\begin{equation*}
    \myVec{\ell}[v] = \log\frac{\tilde{\boldsymbol{c}}[v]}{1-\tilde{\boldsymbol{c}}[v]},
\end{equation*}
and we use the vector $\myVec{\ell}$ to denote the stacking of $\myVec{\ell}[v]$ for all bit indices $v$ corresponding to a given codeword.
 %The hard-decision message, denoted $\hat{\boldsymbol{m}}$, is computed .
%\Nir{The dimensions here are confusing. It looks like $\myVec{\ell}$, and $\myVec{w}$ are vectors, while $\myMat{W}$ is a matrix. This however doesn't settle with how they are used, particularly in \eqref{eq:var_to_check}. My guess is that you use $\myVec{w}[i]$ to denote the $i$th entry of the vector $\myVec{w}$. Is this the case? if so please clarify.}\Tomer{Tried to do so. Hope it is clearer now.}

\begin{figure}
\removelatexerror
  \begin{algorithm}[H]
    \caption{\ac{wbp} Inference}
    \label{alg:wbp}
    \SetAlgoLined
    \SetKwInOut{Input}{Input}
      \Input{\ac{llr} values $\myVec{\ell}$; Parameter vector $\Weights^{\rm WBP}$.} %  \newline 
    \SetKwInOut{Output}{Output}
    \SetKwProg{WBPInference}{\ac{wbp} Inference}{}{}  
    \Output{Soft estimations for message bits $\hat{\boldsymbol{L}}$.}    \WBPInference{}{%$(\boldsymbol{y},\Weights^{\rm SIC})$}{
    
    Initialize $M^{(0)}_{v \rightarrow h} = M^{(0)}_{h \rightarrow v} = 0$.
    
    \For{$q\in\{1,...,Q\}$\label{line:loop}}{ 
            { 
            \For{$(v,h) \in \mathcal{H}$}{
                Update $M^{(q)}_{v \rightarrow h}$ by \eqref{eq:var_to_check}.}\label{stp:wbp_infer1}
            \For{$(v,h) \in \mathcal{H}$}{
                 Update $M^{(q)}_{h \rightarrow v}$ by \eqref{eq:check_to_var}.\label{stp:wbp_infer2}
                }            
            }
        }
    Calculate $\hat{\boldsymbol{L}}$ by \eqref{eq:wbp_final} with $q=Q$.\label{stp:wbp_estimation}
    
    \KwRet{$\hat{\boldsymbol{L}}$} 
    \label{line:return}
  }
  \end{algorithm}
\end{figure}

After initialization, the iterative algorithm begins. Variable-to-check messages are updated according to the rule. i.e.,
\begin{equation} 
\label{eq:var_to_check}
    M^{(q)}_{v \rightarrow h}= \tanh \Big( \frac{1}{2}(\myVec{\ell}[v] +\sum_{\substack{(v,h') \in \mathcal{H} \\ h' \neq h}} \myVec{W}^{(q)}_{h' \rightarrow v}M^{(q-1)}_{h' \rightarrow v}) \Big),
\end{equation}
with the parameters $\Weights^{(q)}=\{\myVec{W}^{(q)}\}$, where $\myVec{W}^{(q)}$ is a weights' matrix. The check-to-variable messages are updated by a non-weighted rule
\begin{equation}
\label{eq:check_to_var}
    M^{(q)}_{h \rightarrow v} = 2\arctanh{ \Big(\prod_{\substack{(v',h) \in \mathcal{H} \\ v' \neq v}} M^{(q)}_{v' \rightarrow h} \Big)}.
\end{equation}
The bit-wise probabilities are obtained from the messages via
\begin{equation}
\label{eq:wbp_final}
    \hat{L}^{(q)}[v]=\tanh \Big(\frac{1}{2}(\myVec{\ell}[v] +\sum_{(v,h')\in \mathcal{H}} \myVec{W}^{(q)}_{h' \rightarrow v}M^{(q-1)}_{h' \rightarrow v})\Big)
\end{equation}

The update rules are performed on every combination of $v,h$ for $Q$ iterations. Finally, the codeword is estimated as the hard-decision on the probability vector $\hat{\myVec{L}}$, which is the stacking of the bit-wise probabilities $\hat{L}^{(Q)}[v]$ for every $v$ in the codeword. The entire inference procedure for a given \ac{wbp} paramterization  $\Weights^{\rm WBP} = \{\Weights^{(q)}\}_{q=1}^Q$ is summarized in Algorithm~\ref{alg:wbp}.

\begin{figure}
\removelatexerror
  \begin{algorithm}[H]
    \caption{Bayesian DeepSIC}
    \label{alg:bayesian_deepsic}
    \SetAlgoLined
    \SetKwInOut{Input}{Input}
    \Input{Channel output $\boldsymbol{y}[i]$; Distribution over  entire architecture $p^{\rm b}(\Weights^{\rm SIC})$; Ensemble size $J$.} %  \newline 
    \SetKwInOut{Output}{Output}
    \SetKwProg{BayesianDeepSICInference}{Bayesian Inference}{}{}
    \Output{Soft estimations for all users $\{\hat{\myVec{P}}_k[i]\}_{k=1}^K$.}    \BayesianDeepSICInference{}{

    \For{$j\in\{1,...,J\}$}{
    Generate $\Weights_j^{\rm SIC} \sim p^{\rm b}(\Weights^{\rm SIC})$;\label{line:generate}

    $\{\hat{\myVec{P}}_{k,j}[i]\}_{k=1}^K \gets$ \textbf{DeepSIC} $(\boldsymbol{y}[i],\Weights_j^{\rm SIC})$;\label{line:estimate}
    }
    \KwRet{$\{\hat{\myVec{P}}_k[i]= \frac{1}{J}\sum_{j=1}^J\hat{\myVec{P}}_{k,j}[i]\}_{k=1}^K$} \label{line:return2}
  }
  \end{algorithm}
\end{figure}

 The parameters of  \ac{wbp},  $\Weights^{\rm WBP}$, can be learned once offline from simulated data and then be re-used throughout the entire transmission since the codebook is static. Specifically, the decoder is trained offline to minimize the following cross-entropy multi-loss \eqref{eq:ce_loss} over the pilots set $\mySet{\Codlen}^{\rm pilot} = \{1,\ldots,r* T^{\rm pilot}\}$  
\begin{equation}
    \label{eq:wbp_loss}
    \mySet{L}_{\rm CE}^{\rm WBP}(\Weights) = \sum_{q=1}^Q \mySet{L}_{\rm CE}^{\rm MWBP}(\Weights^{(q)}),
\end{equation}
with the cross-entropy loss at iteration $q$ defined as
\begin{align}
    & \mySet{L}_{\rm CE}^{\rm MWBP}(\Weights^{(q)}) = \notag \\
   &\quad - \frac{1}{|\mySet{\Codlen}^{\rm pilot}|}\sum_{v\in  \mySet{\Codlen}^{\rm pilot}} \log P_{\Weights^{(q)}}(\myVec{m}[v]|\myVec{\ell}[v], \{ M^{(q-1)}_{h' \rightarrow v}\}).
    \label{eq:module_wbp_loss}
\end{align}
In \eqref{eq:module_wbp_loss}, $P_{\Weights^{(q)}}(\myVec{m}[v]|\myVec{\ell}[v], \{ M^{(q-1)}_{h' \rightarrow v}\})$ is the binary probability vector computed for the $v$th bit of the message, i.e.,  $\myVec{m}[v]$. %\Nir{this should be average over $\Codlen^{\rm pilot}$. Also, where is $P_{\Weights^{(q)}}(\myVec{m}[v]|\myVec{\ell}[v], \{ M^{(q-1)}_{h' \rightarrow v}\})$ defined? Note that you also use this symbol later on so you have to define it }\Tomer{Defined it above.}
%
%\Nir{TODO Tomer - please write the expression explicitly and call it $\mySet{L}_{\rm CE}^{\rm MWBP}$ and refer to it later}\Tomer{Done}. 
 %, whose output is the estimate of the final message $P_{\Weights^{\rm WBP}}(\myVec{m}|\myVec{\ell})$.
	%----------------------------------------------------------------------------------------
	%	Algorithm
	%----------------------------------------------------------------------------------------
	\section{Modular Bayesian Deep Learning}
\label{sec:model_based_bayesian_learning}
%\vspace{-0.1cm} 
Bayesian learning is known to improve the calibration of black-box \acp{dnn} models by preventing overfitting \cite{gal2016dropout,zecchin2023robust}. In this section, we fuse Bayesian learning with hybrid model-based data-driven deep receivers. 
To this end, we first present the direct application of Bayesian learning to deep receivers in Subsection~\ref{subsec:bayesian_learning}, and illustrate the approach for DeepSIC and \ac{wbp} in Subsection~\ref{subsec:naive_bayesian}. Then, we introduce the proposed modular Bayesian framework, and specialize it again to DeepSIC and \ac{wbp} in Subsection~\ref{subsec:model_based_bayesian}. We conclude with a discussion presented in Subsection~\ref{subsec:discussion}.

\subsection{Bayesian Learning}
\label{subsec:bayesian_learning}
Bayesian learning  treats the parameter vector $\Weights$ of a machine learning architecture as a {\em random vector}. This random vector obeys a distribution $p^{\rm b}(\Weights)$ that accounts for the available data and for prior knowledge. Most commonly, %\Nir{if there are other objectives guiding Bayesian learning then replace specifically with "typically" or "the common appproach in Baysian learning"} \Tomer{Fixed}
the distribution is obtained by minimizing the \emph{free energy loss}, also known as negative evidence lower bound (ELBO) \cite{simeone2022machine}, via
\begin{align}
            p^{\rm b}(\Weights) =  \mathop{\arg\min}\limits_{p'} \bigg(& \mathbb{E}_{\Weights \sim p'(\Weights)}[\mySet{L}_{CE}(\Weights)] \notag \\
            &+  \frac{1}{\beta}  \text{KL}(p'(\Weights) || p^{\rm prior}(\Weights)  )\bigg).
            \label{eq:Bayesian_general_sol}
\end{align}
The \ac{kl} divergence term in \eqref{eq:Bayesian_general_sol}, given by 
\begin{equation*}
 \text{KL}\big(p'(\Weights) || p(\Weights)\big)  \triangleq \mathbb{E}_{p'(\Weights)}\Big[\log \big(\frac{p'(\Weights)}{p(\Weights)}\big)\Big],   
\end{equation*}
regularizes the optimization of  $p^{\rm b}(\Weights)$ with a prior distribution $p^{\rm prior}(\Weights)$. The hyperparameter $\beta > 0$,  known as the inverse temperature parameter, balances the regularization  impact \cite{jose2020free,simeone2022machine}. 
 Note that the optimization of \eqref{eq:Bayesian_general_sol}, in which expectation is approximated via empirical averaging, is similar to a regular frequentist loss optimization (see \eqref{eq:frequentist_general_sol}), except for a new \ac{kl} regularization term. Usually, one chooses the prior $p^{\rm prior}(\Weights)$ as a Gaussian, which yields a simple analytical expression~\cite{gal2017concrete} that can be easily calculated and added to the regular loss during \ac{sgd} and its variants.

  Once trained, Bayesian learning aims at having the output conditional distribution approximate the stochastic expectation
\begin{align}
\label{eq:Bayesian_inference}
P_{p^{\rm b}}=\mathbb{E}_{\Weights \sim p^{\rm b}(\Weights)}[ P_{\Weights} ].
\end{align}  
%\Nir{please double check which is conditioned by which. Assuming $x^\text{out}$ is the symbol you wish to decode then it should be the other way around}\Tomer{Switched it.}
The expected value in \eqref{eq:Bayesian_inference} is approximated by  \emph{ensemble} decision, namely, combining multiple predictions, with each decision based on a different realization of $\Weights$ sampled from the optimized distribution $p^{\rm b}(\Weights)$.  Bayesian prediction \eqref{eq:Bayesian_inference} can account for the epistemic uncertainty residing in the optimal model parameters. This way, it can generally obtain a better  calibration performance as compared to frequentist learning~\cite{cohen2022bayesian, zecchin2023robust}.

\begin{figure}
\removelatexerror
  \begin{algorithm}[H]
    \caption{Bayesian \ac{wbp}}
    \label{alg:bayesian_wbp}
    \SetAlgoLined
    \SetKwInOut{Input}{Input}
    \Input{\ac{llr} values $\myVec{\ell}$; Distribution over entire architecture $p^{\rm b}(\Weights^{\rm WBP})$; Ensemble size $J$.} %  \newline 
    \SetKwInOut{Output}{Output}
    \SetKwProg{BayesianWBPInference}{Bayesian Inference}{}{}
    \Output{Soft estimations for message bits $\hat{\boldsymbol{L}}$.}    \BayesianWBPInference{}{%$(\boldsymbol{y},p^{\rm b}(\Weights^{\rm SIC}),J)$}{

    \For{$j\in\{1,...,J\}$}{
    Generate $\Weights_j^{\rm WBP} \sim p^{\rm b}(\Weights^{\rm WBP})$;\label{line:generate}

    $\hat{\boldsymbol{L}}_j \gets$ \textbf{WBP} $(\myVec{\ell},\Weights_j^{\rm WBP})$;\label{line:estimate}
    }
    \KwRet{$\hat{\myVec{L}}= \frac{1}{J}\sum_{j=1}^J\hat{\boldsymbol{L}}_j$} \label{line:return2}
  }
  \end{algorithm}
\end{figure}

\subsection{Conventional Bayesian Learning for Deep Receivers}
\label{subsec:naive_bayesian}
\vspace{-0.1cm}

The conventional application of Bayesian learning includes obtaining the distribution over the set of model parameters vectors $p^{\rm b}(\Weights)$. Afterwards, one can proceed to inference,  where an ensembling over multiple realizations of parameters vectors, drawn from $p^{\rm b}(\Weights)$, leads to more reliable estimate. Below, we specialize this case for our main running examples of deep receiver architectures: DeepSIC and \ac{wbp}.

\subsubsection{DeepSIC}
\label{subsubsec:naive_bayesian_deepsic}
The DeepSIC equalizer is formulated in Algorithm~\ref{alg:deepsic} for a deterministic (frequentist) parameter vector. When conventional Bayesian learning is employed, one obtains a distribution $p^{\rm b}(\Weights^{\rm SIC})$ over the set of model parameter vectors $\Weights^{\rm SIC}$ by minimizing the free energy loss \eqref{eq:Bayesian_general_sol}. Using these distribution, the prediction for information symbols is computed as
\begin{align}
\label{eq:baysian_deepsic}
    &P_{p^{\rm b}}(\myVec{s}_k[i]|\myVec{y}[i])= \mathbb{E}_{\Weights^{\rm SIC} \sim p^{\rm b}(\Weights^{\rm SIC})}   \big[  P_{\Weights^{\rm SIC}}(\myVec{s}_k[i]|\myVec{y}[i])   \big] ,
\end{align}
for each data symbol at $i \in \mySet{\Blklen}^{\rm info}$,
with soft prediction  $\hat{\myVec{P}}_k$ following \eqref{eq:soft_freq_out}. The resulting inference procedure is described in Algorithm~\ref{alg:bayesian_deepsic}.

\begin{figure}[t!]
\removelatexerror
  \begin{algorithm}[H]
    \caption{Modular Bayesian DeepSIC}
    \label{alg:model_based_bayesian_deepsic}
    \SetAlgoLined
    \SetKwInOut{Input}{Input}
    \Input{Channel output $\boldsymbol{y}[i]$; Distributions per module $\{\{ p^{\rm mb}(\Weights^{(k,q)}) \}_{k=1}^K\}_{q=1}^Q$; Ensemble size $J$.} %  \newline 
    \SetKwInOut{Output}{Output}
    \SetKwProg{ModularBayesianDeepSICInference}{Modular Bayesian Inference}{}{}
    \Output{Soft estimations for all users $\{\hat{\myVec{P}}_k[i]\}_{k=1}^K$.}    \ModularBayesianDeepSICInference{}{%$(\boldsymbol{y},\{\{ p^{\rm mb}(\Weights^{(k,q)}) \}_{k=1}^K\}_{q=1}^Q,J)$}{

    Generate an initial guess $\{\hat{\myVec{P}}^{(k,0)}\}_{k=1}^K$.
    
    \For{$q\in\{1,...,Q\}$\label{line:loop2}}{ 
            \For{$k\in\{1,...,K\}$}{ 
                \texttt{Bayesian Estimation Part}\
                
                \For{$j\in\{1,...,J\}$}{
                Generate $\Weights^{(k,q)}_j \sim p^{\rm mb}(\Weights^{(k,q)})$\; \label{line:generate2} 

                Calculate $\hat{\myVec{P}}_j^{(k,q)}$ by \eqref{eq:soft_freq_out} using $\Weights^{(k,q)}_j$.
                }

                 $\hat{\myVec{P}}^{(k,q)}=\frac{1}{J}\sum_{j=1}^J \hat{\myVec{P}}_j^{(k,q)}{} $\;
            }            
        }
    \KwRet{$\{\hat{\myVec{P}}_k[i]=\hat{\myVec{P}}^{(k,Q)}\}_{k=1}^K$} \label{line:return3}
  }
  \end{algorithm}
\end{figure}

While the final prediction $P_{\Weights^{\rm SIC}}(\myVec{s}_k[i]|\myVec{y}[i])$ is obtained by ensembling the predictions of multiple parameter vectors $\Weights^{\rm SIC} \sim p^{\rm b}(\Weights^{\rm SIC})$, the intermediate predictions made during SIC iterations, i.e., $\hat{\myVec{P}}^{(k,q)}$ with $q < Q$, are not combined to approach a stochastic expectation. 
Thus, this form of Bayesian learning only encourages the outputs, i.e., the predictions at iteration $Q$, to be well calibrated, not accounting for the internal modular structure of the architecture. The soft estimates learned by the internal modules may still be overconfident, propagating inaccurate decisions to subsequent modules. 

\subsubsection{WBP}
\label{subsubsec:naive_bayesian_deepsic} 
Bayesian learning of \ac{wbp} obtains a distribution $p^{\rm b}(\Weights^{\rm WBP})$  by the minimization of the free energy loss \eqref{eq:Bayesian_general_sol} using the pilots set $\mySet{\Bitslen}^{\rm pilot}$.  The training occurs once, offline. With the trained architecture, the message bits are decoded via 
\begin{align}
\label{eq:baysian_wbp}
    &\! P_{p^{\rm b}}(\myVec{m}[v]|\myVec{\ell}[v]) \!=  \! \mathbb{E}_{\Weights^{\rm WBP} \sim p^{\rm b}(\Weights^{\rm WBP})}   \big[  P_{\Weights^{\rm WBP}}(\myVec{m}[v]|\myVec{\ell}[v])   \big] ,
\end{align}
for each data bit of index $v$ in the information bits. 
The resulting inference procedure is described in Algorithm~\ref{alg:bayesian_wbp}.  As noted for DeepSIC, the resulting formulation only considers the soft estimate representation taken at the final iteration of index $Q$ when approaching the stochastic expectation. Doing so fails to leverage the modular and interpretable operation of \ac{wbp} to boost calibration of its internal message.

\subsection{Modular Bayesian Learning for Deep Receivers}
\label{subsec:model_based_bayesian}
The conventional application of Bayesian deep learning is concerned with calibrating the final output of a \ac{dnn}. As such, it does not exploit the modularity of model-based data-driven architectures, and particularly the ability to relate their internal features into probability measures.
Here, we extend Bayesian deep learning to leverage this architecture of modular deep receivers. This is achieved by calibrating the intermediate blocks via Bayesian learning, combined with a dedicated {\em sequential Bayesian learning} procedure, as detailed next for DeepSIC and WBP. 

\subsubsection{DeepSIC}
\label{subsubsec:modular_bayesian_deepsic} 
During online training, we optimize the distribution $p^{\rm mb}(\Weights^{(k,q)})$ over the parameter vector $\Weights^{(k,q)}$ of each  block of index $(k,q)$, i.e.,  $\Weights^{(k,q)}\sim p^{\rm mb}(\Weights^{(k,q)})$, individually. These distributions are  obtained by minimizing the free energy loss on a per-module basis as
\begin{align}
    p^{\rm mb}(\Weights^{(k,q)}) =  &\mathop{\arg\min}\limits_{p'} \bigg( \mathbb{E}_{\Weights^{(k,q)} \sim p'(\Weights^{(k,q)})}[\mySet{L}_{CE}^{\rm  MSIC}(\Weights^{(k,q)})] \notag \\ &\quad + \frac{1}{\beta}  \text{KL}(p'(\Weights^{(k,q)}) || p^{\rm prior}(\Weights^{(k,q)})  )\bigg),
    \label{eq:model_based_Bayesian_solution} 
\end{align}
with the module-wise loss $\mySet{L}_{CE}^{\rm  MSIC}(\Weights^{(k,q)})$ defined in \eqref{eq:module_deepsic_loss}. 

The core difference in the formulation of the training objective in \eqref{eq:model_based_Bayesian_solution}  as compared to \eqref{eq:baysian_deepsic} is that a single module is considered, rather than the entire model. Accordingly, the learning is done in a {\em sequential} fashion, where we first use \eqref{eq:model_based_Bayesian_solution} to train the modules corresponding to iteration $q=1$, followed by training the modules for $q=2$, and so on. This sequential Bayesian learning utilizes the calibrated soft estimates computed by using the data set at the $q$th iteration as inputs when training the modules at iteration $q+1$. Doing so does not only improves the calibration of the intermediate modules, but also encourages them to learn with better calibrated soft estimates of the interfering symbols. 

\begin{figure}
\removelatexerror
  \begin{algorithm}[H]
    \caption{Modular Bayesian \ac{wbp}}
    \label{alg:model_based_bayesian_wbp}
    \SetAlgoLined
    \SetKwInOut{Input}{Input}
    \Input{\ac{llr} values $\myVec{\ell}$; Distributions per module $\{ p^{\rm mb}(\Weights^{(q)}) \}_{q=1}^Q$; Ensemble size $J$.} %  \newline 
    \SetKwInOut{Output}{Output}
    \SetKwProg{ModularBayesianWBPInference}{Modular Bayesian Inference}{}{}
    \Output{Soft estimations for message bits $\hat{\myVec{L}}$.}    \ModularBayesianWBPInference{}{%$(\boldsymbol{y},\Weights^{\rm SIC})$}{
    
    Initialize $M^{(q)}_{v \rightarrow h} = M^{(q)}_{h \rightarrow v} = 0$.
    
    \For{$q\in\{1,...,Q\}$\label{line:loop}}{ 
            { 
            \For{$(v,h) \in \mathcal{H}$}{
                
                \texttt{Bayesian Estimation Part}\
                
                \For{$j\in\{1,...,J\}$}{
                    
                    Generate $\Weights^{(q)}_j \sim p^{\rm mb}(\Weights^{(q)})$\;
                    Compute $M^{(q)}_{v \rightarrow h,j}$ by \eqref{eq:var_to_check} using $\Weights^{(q)}_j$\;
                    }
                $M^{(q)}_{v \rightarrow h}=\frac{1}{J}\sum_{j=1}^J M^{(q)}_{v \rightarrow h,j} $
                }
            \For{$(v,h) \in \mathcal{H}$}{
                 Update $M^{(q)}_{h \rightarrow v}$ by \eqref{eq:check_to_var}.\label{stp:wbp_infer2}
                }           
            }
        }
    Calculate $\hat{\myVec{L}}$ by \eqref{eq:wbp_final} with $q=Q$.\label{stp:wbp_estimation}
    
    \KwRet{$\hat{\myVec{L}}$} 
    \label{line:return}
  }
  \end{algorithm}
\end{figure}

% In \eqref{eq:model_based_Bayesian_solution}  the model-based cross-entropy loss function for the parameter vector $\Weights^{(k,q)}$ is defined as 
%\Nir{the notations are inconsistent (MBSIC, MSIC?). Also, this should be defined in the frequentist DeepSIC, not here}\Tomer{Done.}
% \begin{align}
%     \label{eq:model_based_deepSIC_ce_loss}
%     \begin{split}
%     &\mySet{L}^{\rm  MBSIC}_{\rm CE}(\Weights^{(k,q)}) = \\ &-\sum_{i \in \mySet{B^{\rm pilot}}} \log P_{\Weights^{(k,q)}}( \myVec{s}_k[i]|\myVec{y}[i], \{ \hat{P}^{(l,q-1)}[i] \}_{l\neq k}).
%     \end{split}
% \end{align}
Once the modules corresponding to the $q$th iteration are trained, the soft estimate $\hat{\myVec{P}}^{(k,q)}[i]$ for the  $k$th symbol at the $q$th iteration is computed by taking into account the uncertainty of the corresponding model parameter vector $\Weights^{(k,q)}$. The resulting expression is given by
\begin{align*} 
   \hat{\myVec{P}}^{(k,q)}[i] = \Big\{ &\mathbb{E}_{\Weights^{(k,q)} \sim p^{\rm mb}(\Weights^{(k,q)})  }\\ &\quad   \left[P_{\Weights^{(k,q)}}(\myVec{s}|\myVec{y}[i], \{\hat{\myVec{P}}^{(l,q-1)}[i]\}_{l\neq k})\right]\Big\}_{\myVec{s}\in \mySet{S}}, 
\end{align*}
where again the stochastic expectation is approximated via ensembling.

Having computed a possibly different distribution for the weights of each intermediate module, prediction of the  information symbols is  computed by setting $\hat{\myVec{P}}_k[i]= \hat{\myVec{P}}^{(k,Q)}[i]$
for each data symbol at $i \in \mySet{\Blklen}^{\rm info}$.
 The resulting  inference algorithm for the trained modular Bayesian DeepSIC is summarized in Algorithm~\ref{alg:model_based_bayesian_deepsic}.

 \subsubsection{Weighted Belief Propagation}
\label{subsubsec:modular_bayesian_wbp}
%\Nir{Please formulate this. It can be brief but still explain the learning and inference. Also explain why the internal features are viewed as probablity measures that need to be calibrated.} \Tomer{Done.}
Each internal messages of  \ac{bp} represents the belief on the value of a variable or check node. Modelling each message exchange iteration via a Bayesian deep architecture enables more reliable information exchange of such belief values. By doing so,  one can prevent undesired quick convergence of the beliefs to polarized values from overconfident estimates, thus enhancing the reliability of the final estimate. 

Carrying out \ac{wbp} via modular Bayesian deep learning requires learning a {\em per-module} basis distribution. Specifically, the per-module distribution is obtained  (via pre-training) by minimizing the free energy loss per module, i.e., each module $q$ solves
\begin{align*}
    p^{\rm mb}(\Weights^{(q)}) =  &\mathop{\arg\min}\limits_{p'} \bigg( \mathbb{E}_{\Weights^{(q)} \sim p'(\Weights^{(q)})}[\mySet{L}_{CE}^{\rm  MWBP}(\Weights^{(q)})] \notag \\ &\quad + \frac{1}{\beta}  \text{KL}(p'(\Weights^{(q)}) || p^{\rm prior}(\Weights^{(q)})  )\bigg), 
\end{align*}
where the module-wise loss $\mySet{L}_{CE}^{\rm  MWBP}(\Weights^{(q)})$ is defined in \eqref{eq:module_wbp_loss}. As detailed for modular Bayesian DeepSIC, training is carried out via sequential Bayesian learning, gradually learning the distribution of the parameters of each iteration, while using its calibrated outputs when training the parameters of the subsequent iteration.

In particular, each intermediate probability value is calculated based on realizations of parameters drawn from the obtained distribution, i.e.,
\begin{align*} 
    &\hat{L}^{(q)}[v] = \\ &\qquad \mathbb{E}_{\Weights^{(q)} \sim p^{\rm mb}(\Weights^{(q)})  }  \left[P_{\Weights^{(q)}}(\myVec{m}[v]|\myVec{\ell}[v], \{ M^{(q-1)}_{h' \rightarrow v}\} \right],
\end{align*}
where ensembling is used to approximate the expected value.
%\Nir{you have $[i]$ on one side and $v$ on the other.... You have to carefully proofread the math here and make sure it is consistent} \Tomer{Thanks for spotting it out, the other variables are related. This is simply the alteration of the equation from the DeepSIC part.}
The proposed inference procedure for modular Bayesian \ac{wbp} is described in Algorithm~\ref{alg:model_based_bayesian_wbp}. 

% TODO NIR CONTINUE FROM HERE

\subsection{Discussion}
\label{subsec:discussion}

The core novelty of the algorithms formulated above lies in the introduction of the concept of modular Bayesian deep learning, specifically designed for hybrid model-based data-driven receivers, such as DeepSIC and \ac{wbp}. Unlike conventional Bayesian approaches (see Algorithms~\ref{alg:bayesian_deepsic} and \ref{alg:bayesian_wbp}) that focus solely on the output, our method applies Bayesian learning to each intermediate module, leading to increased reliability and performance at every stage of the network. Algorithms~\ref{alg:model_based_bayesian_deepsic} and \ref{alg:model_based_bayesian_wbp} illustrate the Bayesian estimation in intermediate layers. 

Bayesian learning involves adding a loss term to the frequentist loss optimization, which in turn is optimized using conventional first-order methods, e.g., \ac{sgd} and its variants. Accordingly, compared to training of standard frequentist architectures, it involves a negligible increase in  the training optimization overhead. However, the inference complexity increases by the ensemble factor, as one has to run inference multiple time, but this could be parallelized via a dedicated Hardware. As such, standard Bayesian learning and our modular Bayesian variation have similar overall complexity overhead, while our scheme propagates more reliable decisions throughout the model, resulting in higher performance and reliability to classification tasks, such as detection and decoding. The consistent benefits of modular Bayesian learning are demonstrated through simulation studies in Section~\ref{sec:Simulation}.

Our training algorithm relies on a sequential operation which gradually learns the distribution of each iteration-based module in a deep unfolded architecture, while using the calibrated predictions of preceding iterations. This learning approach has three core gains: $(i)$ it facilitates individually assessing each module based on its probabilistic prediction; $(ii)$ it enables the trainable parameters of subsequent iterations to adapt to already calibrated predictions of their preceding modules; and $(iii)$ it facilitates an efficient usage of the available limited data obtained from pilots, as it reuses the entire data set for training each module. Specifically, when training each module individually, the data set is used to adapt a smaller number of parameters as compared to jointly training the entire architecture as in conventional end-to-end training. Nonetheless, this form of learning is expected to be more lengthy as compared to conventional end-to-end frequentist learning, due to its sequential operation combined with the excessive computations of Bayesian learning \eqref{eq:Bayesian_general_sol}. While one can possibly reduce training latency via, e.g., optimizing the learning algorithm~\cite{raviv2022online}, such extensions of our framework are left for future research.

	\section{Numerical Evaluations}
\label{sec:Simulation}
%% \vspace{-0.1cm} 
In this section we numerically quantify the benefits of the proposed modular Bayesian deep learning framework as compared to both frequentist learning and conventional Bayesian learning. We focus on uplink \ac{mimo} communications with  DeepSIC detection and \ac{wbp} decoding. We note that the relative performance of DeepSIC and \ac{wbp} with frequentist learning in comparison to state-of-the-art receiver methods was studied in \cite{shlezinger2019deepSIC} and \cite{nachmani2018deep}, respectively. Accordingly, here we focus on evaluating the performance gains of a receiver architecture based on DeepSIC and \ac{wbp} {\em when modular Bayesian learning is adopted}, using as benchmarks frequentist learning and standard, non-modular, Bayesian learning.  Note that DeepSIC and its variants are adapted online, while \ac{wbp} and its variants are trained once offline as the codebook does not change throughout transmission.
Specifically, after formulating the experimental setting and methods in Subsection~\ref{subsec:compared_approaches}, we study the effect of modular Bayesian learning for equalization and detection separately in Subsections~\ref{subsec:detection_study}-\ref{subsec:decoding_performance}, respectively. Then, we asses the calibration of modular Bayesian equalization and its contribution to downstream soft decoding task in Subsection~\ref{subsec:gains_to_downstream_tasks}, after the full modular Bayesian receiver is evaluated in Subsection~\ref{subsec:final_comparison}.

 \begin{figure*}[t]
    \centering
    \begin{subfigure}[a]{0.48\textwidth}  
        \centering 
        \includegraphics[width = 0.9\columnwidth]{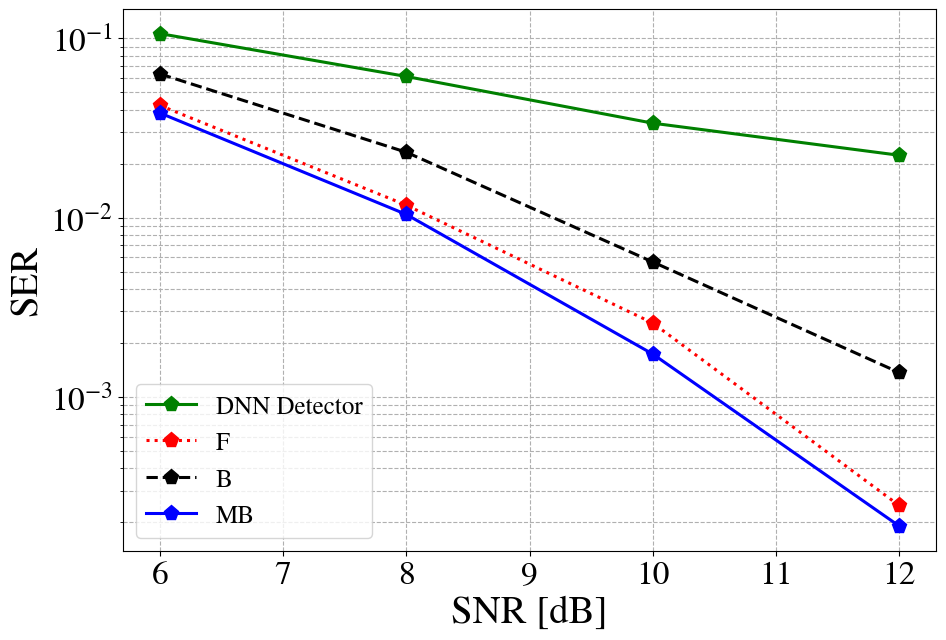}
	\caption{QPSK constellation}
    \end{subfigure}
    \hfill
    \begin{subfigure}[a]{0.48\textwidth}  
        \centering 
        \includegraphics[width = 0.9\columnwidth]{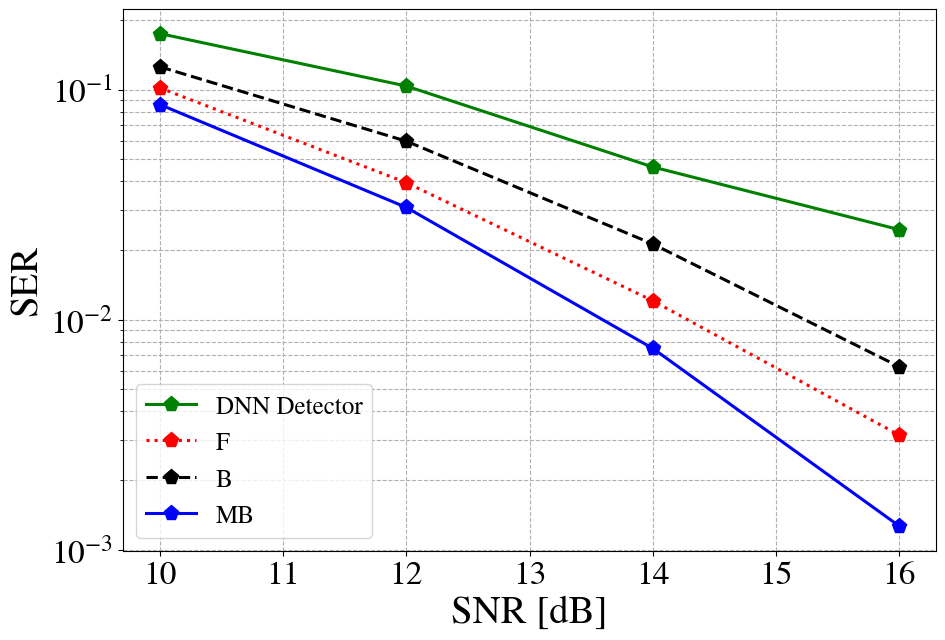}
	\caption{8-PSK constellation}
    \end{subfigure}
    \caption{Static linear synthetic channel: \acs{ser} as a function of the SNR for different training methods. At each block, training is done with $384$ pilots,  and the \acs{ser} is computed on $14,976$ information symbols from the QPSK or 8-PSK constellation. Results are averaged over $10$ blocks  per point.}
    \label{fig:mimo_ser_linear_results}
    \vspace{-0.4cm}
\end{figure*}

 \begin{figure*}[t]
    \centering
    \begin{subfigure}[a]{0.48\textwidth}  
        \centering 
        \includegraphics[width = 0.9\columnwidth]{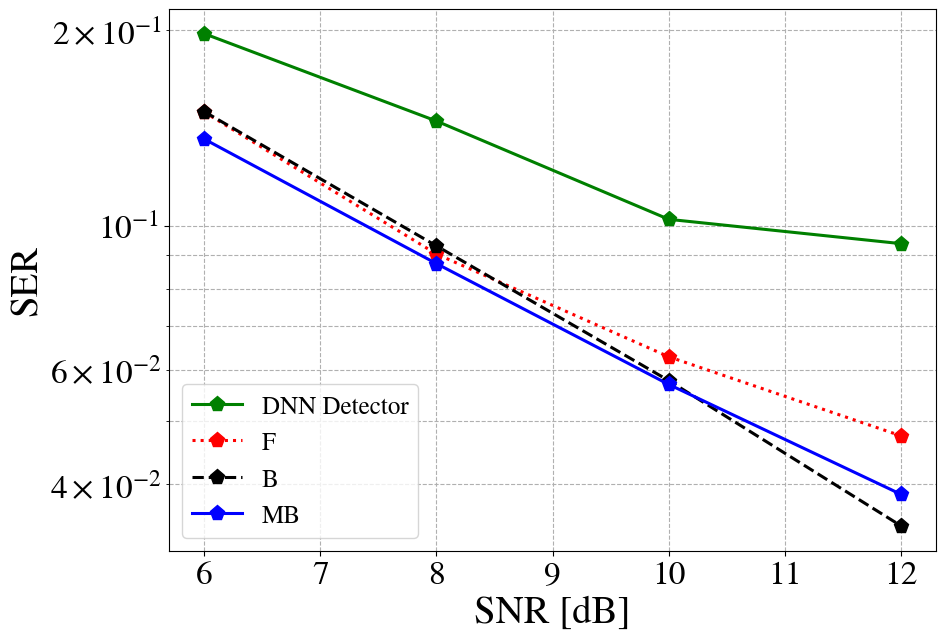}
	\caption{QPSK constellation}
    \end{subfigure}
    \hfill
    \begin{subfigure}[a]{0.48\textwidth}  
        \centering 
        \includegraphics[width = 0.9\columnwidth]{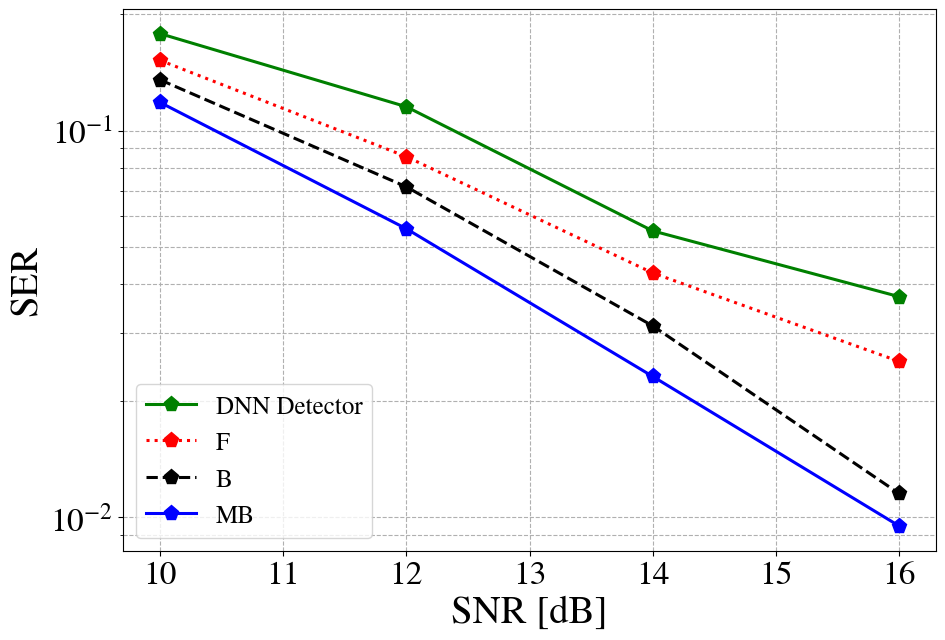}
	\caption{8-PSK constellation}
    \end{subfigure}
    \caption{Static non-linear synthetic channel: \acs{ser} as a function of the SNR for different training methods. At each block, training is done with $384$ pilots,  and the \acs{ser} is computed on $14,976$ information symbols from the QPSK or 8-PSK constellation. Results are averaged over $10$ blocks  per point.}
    \label{fig:mimo_ser_non_linear_results}
    \vspace{-0.4cm}
\end{figure*}

%-----------------------------------
%	Evaluated Equalizers
%-----------------------------------
% \vspace{-0.2cm}
\subsection{Architecture and Learning Methods}
\label{subsec:compared_approaches}
{\bf Architectures:}
The model-based \ac{dnn}-aided equalizer used in this study is based on the DeepSIC architecture \cite{shlezinger2019deepSIC}, with \ac{dnn}-modules comprised of two \ac{fc} layers with a hidden layer size of $16$ neurons and ReLU activations unrolled over $Q=3$ iterations. We also employ a black-box fully-connected \ac{dnn} as a benchmark. This architecture is composed of four \ac{fc} layers with hidden layer size of 32 and ReLU activations in-between.
The decoder chosen is based on \ac{wbp}~\cite{nachmani2016learning}, which is implemented over $Q=5$ iterations with learnable parameters corresponding  to messages of the variable-to-check layers \eqref{eq:var_to_check}. Choosing $Q=5$ layers was empirically observed to reach satisfactory high performance gain over the vanilla \ac{bp} \cite{nachmani2016learning,nachmani2018deep}. 

\smallskip
{\bf Learning Methods:}
For both detection and decoding, we compare the proposed modular Bayesian learning to both frequentist learning and  Bayesian learning. 
Specifically, we employ the following three learning schemes:
\begin{itemize}
    \item {\em Frequentist (F):} This method, labelled as ``F" in the figures, applies the conventional learning for training DeepSIC and \ac{wbp} as detailed in Subsection~\ref{subsec:frequentist}.   
    \item {\em Bayesian (B):} The conventional Bayesian receiver, referred to as ``B", carries out optimization of the distribution $p^{\rm b}(\Weights)$ in an end-to-end fashion as described in Subsection~\ref{subsec:naive_bayesian}.
    \item {\em Modular Bayesian (MB):} The proposed modular Bayesian learning scheme, introduced in Subsection~\ref{subsec:model_based_bayesian}, 
 that applies  Bayesian learning separately to each module, and is labelled as ``MB" in the figures.
\end{itemize}

For the Bayesian learning schemes ``B" and ``MB", we adopt the Monte Carlo (MC) dropout scheme \cite{gal2016dropout, boluki2020learnable},  which parameterizes the distribution $p^{\rm b}(\Weights)$ in terms of a nominal model parameter vector  and dropout logits vector $\myVec{\alpha} \in \mySet{R}^{|\Weights|}$. 
 The dropout logits vector is obtained via cross-entropy loss minimization ~\cite[equation (7)]{boluki2020learnable}, and is added the frequentist loss and the \ac{kl} term. Assuming Gaussian prior, one obtains the \ac{kl} term specified in \cite[Eq. (4)]{gal2016dropout}. During Bayesian training, we set the inverse temperature parameter $\beta = 10^{4}$; and during Bayesian inference we set the ensemble size as $J=5$ for DeepSIC and as $J=3$ for \ac{wbp}. For all of DeepSIC variations, optimization is done via Adam with $500$ iterations and learning rate of $5\cdot 10^{-3}$. For \ac{wbp}, Adam with $500$ iterations and learning rate of $10^{-3}$ was used.  
These values were set empirically to ensure convergence\footnote{The source code used in our experiments is available at \hyperlink{https://github.com/tomerraviv95/bayesian-learning-in-receivers-for-decoding}{https://github.com/tomerraviv95/bayesian-learning-in-receivers-for-decoding}}.

%-----------------------------------
%	Detection Study
%-----------------------------------
% \vspace{-0.2cm}
\subsection{Detection Performance}
\label{subsec:detection_study}

We begin by evaluating the proposed modular Bayesian deep learning framework for symbol detection using the DeepSIC equalizer.  We consider  two types of different channels: $(\emph{i})$ a static synthetic linear channel, and $(\emph{ii})$  a time-varying channel obtained from the physically compliant COST simulator~\cite{liu2012cost}, both with additive Gaussian noise. The purpose of this study is to illustrate the benefits of employing the modular Bayesian approach to improve the detection performance in terms of \ac{ser}.  Moreover, we compare model-based deep learning architectures  to black-box \acp{dnn}, contrasting all the training variants of DeepSIC with the black-box fully-connected \ac{dnn}, that is trained with the frequentist method. We refer to this approach as {\em DNN Detector} in the figures.

\subsubsection{Static Linear Synthetic Channels}
\label{subsubsec:static_linear}

The considered input-output relationship of the memoryless Gaussian \ac{mimo} channel is given by
\begin{equation}
\label{eqn:GaussianMIMO}
\myVec{y}[i] = \myMat{H}\myVec{s}[i] + \myVec{w}[i],
\end{equation}
where $\myMat{H}$ is a fixed $N\times K$ channel matrix, and $\myVec{w}[i]$ is a complex white Gaussian noise vector.  The  matrix $\myMat{H}$ models spatial exponential received power decay, and its entries are given by
$\left( \myMat{H}\right)_{n,k} = e^{-|n-k|}$, for each $n \in \{1,\ldots, N\}$ and $ k \in \{1,\ldots, K\}$. The pilots and the transmitted symbols are generated in a uniform i.i.d. manner over two different constellations, the first being the \ac{qpsk} constellation, i.e., $\mathcal{S}=\{(\pm \rfrac{1}{\sqrt{2}},\pm \rfrac{1}{\sqrt{2}})\}$, and the second being 8-ary phase shift keying (8-PSK) constellation, i.e., $\mathcal{S}=\{e^{\frac{j\pi\ell}{4}}| \ell \in \{0, \ldots, 7\}\}$. For both cases, the number of users and antennas are set as $K=N=4$.

\begin{figure*}[t]
    \centering
    \begin{subfigure}[a]{0.48\textwidth}  
        \centering 
        \includegraphics[width = 0.9\columnwidth]{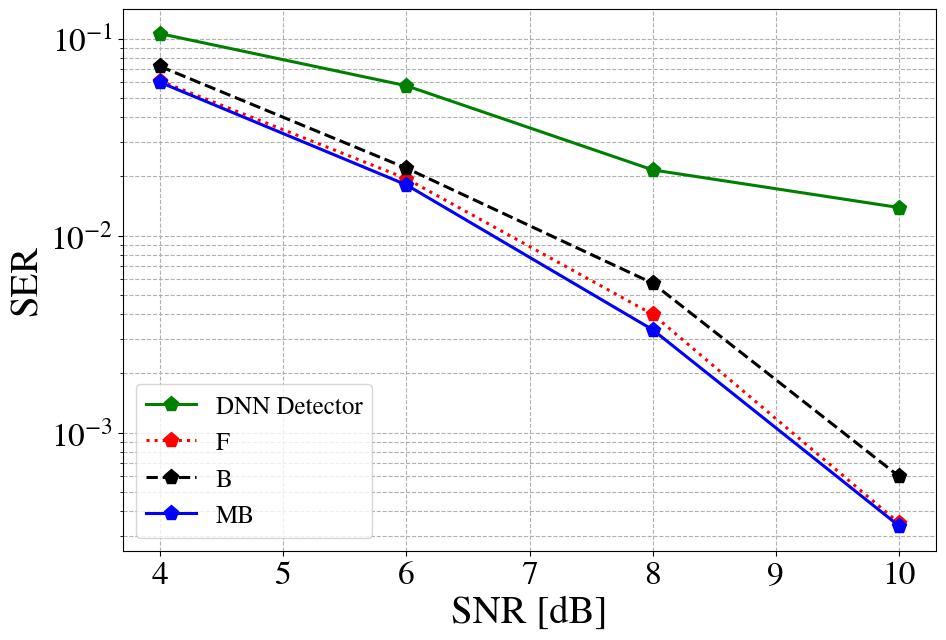}
	\caption{QPSK constellation}
    \end{subfigure}
    \hfill
    \begin{subfigure}[a]{0.48\textwidth}  
        \centering 
        \includegraphics[width = 0.9\columnwidth]{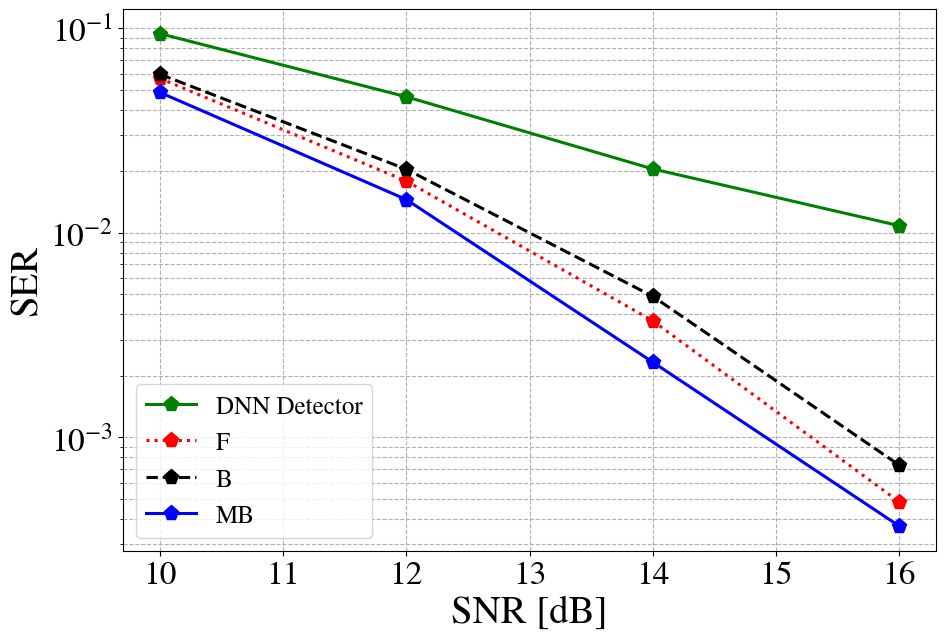}
	\caption{8-PSK constellation}
    \end{subfigure}
    \caption{{Time-varying COST channel:} \acs{ser} as a function of the SNR for different training methods. At each block, training is done with $384$ pilots, and the \acs{ser} is computed on $14,976$ information symbols from the QPSK or 8-PSK constellation. Results are averaged over $10$ blocks  per point.}
    \label{fig:mimo_ser_static_results}
    % \vspace{-0.4cm}
\end{figure*}

We evaluate the accuracy of hard detection in terms of \ac{ser}, i.e., as the average $\frac{1}{K}\sum_k\mathbb{E}[\mathbbm{1}(\hat{\myVec{s}}_k[i]\neq\myVec{s}_k[i])]$, with $\mathbbm{1}(\cdot)$ being the indicator function, and where the expectation is estimated using the transmitted symbols and the noise in \eqref{eqn:GaussianMIMO}. 
 Specifically, for the \ac{qpsk} (8-PSK) constellation, we adopt  $\Blklen^{\rm pilot}= 128$ ($\Blklen^{\rm pilot}= 384$) pilot symbols and $\Blklen^{\rm info} =15,232$ ($\Blklen^{\rm info} =14,976$) information symbols, for a total number of symbols $\Blklen^{\rm tran}= 15,360$ in a single block for both constellations.

 The resulting \ac{ser} performance versus \ac{snr} are reported in  Fig.~\ref{fig:mimo_ser_linear_results}. The gains of the modular Bayesian DeepSIC (``MB") are apparent for both constellations. Specifically, ``MB" consistently outperforms the frequentist DeepSIC  (``F") by up to $0.2$ dB under \ac{qpsk}, and by up to $1$ dB under 8-PSK in medium-to-high \acp{snr}. The Bayesian (``B'') method falls off from both the frequentist and the modular Bayesian approaches.
The explanation for the performance gap lies in the severe data-deficit conditions. Due to limited data, the frequentist and the black-box Bayesian (``B") methods are severely overfitted, specifically in  the 8-PSK case. In particular, conventional Bayesian learning succeeds in  calibrating the output of the receiver, but not of the intermediate blocks, increasing the reliability at the expense of the accuracy.  Also, the \ac{dnn} detector is observed to be inferior to DeepSIC, hinting to the importance of relying on model-based architecture in the small labelled-data regime. Since it is notably outperformed by the frequentist DeepSIC, we do not include Bayesian learning variants of this approach.

It is also noted that the gain of the proposed ``MB'' approach as compared to ``F'' and ``B'' becomes more pronounced with an increasing \ac{snr} level. To understand this behavior, it is useful to decompose the uncertainty that affects data-driven solutions into \emph{aleatoric uncertainty} and \emph{epistemic uncertainty}. Aleatoric uncertainty is the inherent uncertainty present in data, e.g., due to channel noise; while epistemic uncertainty is the uncertainty that arises from lack of training data. Since both the frequentist and Bayesian learning suffer equally from aleatoric uncertainty, the gain of Bayesian learning becomes highlighted with reduced aleatoric uncertainty, i.e., with an increased \ac{snr} (see, e.g., \cite[Sec. 4.5.2]{simeone2022machine}  for a details).

\subsubsection{Static Non-Linear Synthetic Channels}
\label{subsubsec:non_static_linear}

  Next, we also evaluate the methods on a non-linear variant of the previous channel.The considered input-output relationship of the memoryless Gaussian \ac{mimo} channel is given by
\begin{equation}
\label{eqn:GaussianMIMONonLinear}
\myVec{y}[i] = \tanh{\Big(\frac{1}{2} \myMat{H}\myVec{s}[i] + \myVec{w}[i]\Big)}.
\end{equation}
 This operation may represent, e.g., non-linearities induced by the receiver acquisition hardware, as the hyperbolic tangent function is commonly employed to model the saturation effect observed at the receiver due to the interplay of resistance, capacitance, and circuit sensitivity. Specifically, the model in \eqref{eqn:GaussianMIMONonLinear} can approximately describe the behaviour of non-linear power amplifiers, that are characterized by a linear response near zero that becomes saturated as the input deviates from this point \cite{clerckx2018fundamentals}.

We employ again $K=N=4$, with the rest of the hyperparameters set similar to Subsection~\ref{subsubsec:static_linear}. We report in Fig~\ref{fig:mimo_ser_non_linear_results} the \ac{ser} as a function of the \ac{snr} for both \ac{qpsk} and 8-PSK after the transmission of 10 blocks. It can be observed from the figure that the proposed ``MB'' method surpasses all other methods across most \acp{snr}, with now the ``B'' approach achieving better results than the ``F'' method. The \ac{dnn} detector still suffers from lack of training data as compared to its number of trainable parameters, resulting in the worst performance.

\subsubsection{Time-Varying Linear COST Channels}
\label{subsubsec:cost_detection}
Finally, we evaluate the proposed model-based Bayesian scheme under the time-varying COST 2100  geometry-based stochastic channel model \cite{liu2012cost}.  The simulated setting represents multiple users moving in an indoor setup while switching between different microcells. Succeeding on this scenario requires high adaptivity, since there is considerable variability in the channels observed in different blocks. 
The employed channel model has the channel coefficients change in a block-wise manner. The  instantaneous channel matrix is simulated using the wide-band indoor hall $5$ GHz setting of COST 2100 with single-antenna elements, which  creates $4 \times 4 = 16$ narrow-band channels when setting $K=N=4$.

For the same hyperparameters as in Subsection~\ref{subsubsec:static_linear}, in Fig.~\ref{fig:mimo_ser_static_results}, we report the \ac{ser} as a function of the \ac{snr} for both \ac{qpsk} and 8-PSK after the transmission of 10 blocks. The main conclusion highlighted above, in the synthetic channel, is confirmed in this more realistic setting. It is systematically observed that modular Bayesian learning yields notable \ac{ser} improvements, particularly for higher-order constellations, where overfitting is more pronounced.  The \ac{dnn} detector's performance is yet far from the performance of the other approaches, thus we omit it in all the next simulations.

%-----------------------------------
%	Gains To Downstream Tasks
%-----------------------------------
%\vspace{-0.2cm}
\subsection{Decoding Performance}
\label{subsec:decoding_performance}
We proceed to evaluating the contribution of modular Bayesian learning to channel decoding. To that end, prior to the symbols' transmission, the message is encoded via polar code $(128,64)$ (as in the 5G standard).  The \ac{wbp} decoder is either trained using frequentist (``F"), Bayesian (``B''),  or modular Bayesian (``MB'') methods, and is applied following soft equalization using DeepSIC. We henceforth use the notation ``X/Y'' with $\text{X,Y}\in\{\text{F, B, MB}\}$  to indicate that the soft equalizer whose outputs are fed to \ac{wbp} for decoding is DeepSIC  trained using method ``X'', while \ac{wbp} is trained using ``Y''. As our focus in this subsection is only on the decoder, we consider frequentist detection only.  In all cases, we trained DeepSIC, and then trained the \ac{wbp} decoder on the soft outputs obtained by the corresponding DeepSIC detector according to \eqref{eq:wbp_loss}. This way, the decoder is optimized to make use of the detector's soft outputs. 

\begin{figure}
	\centering
	\includegraphics[width = 0.9\columnwidth]{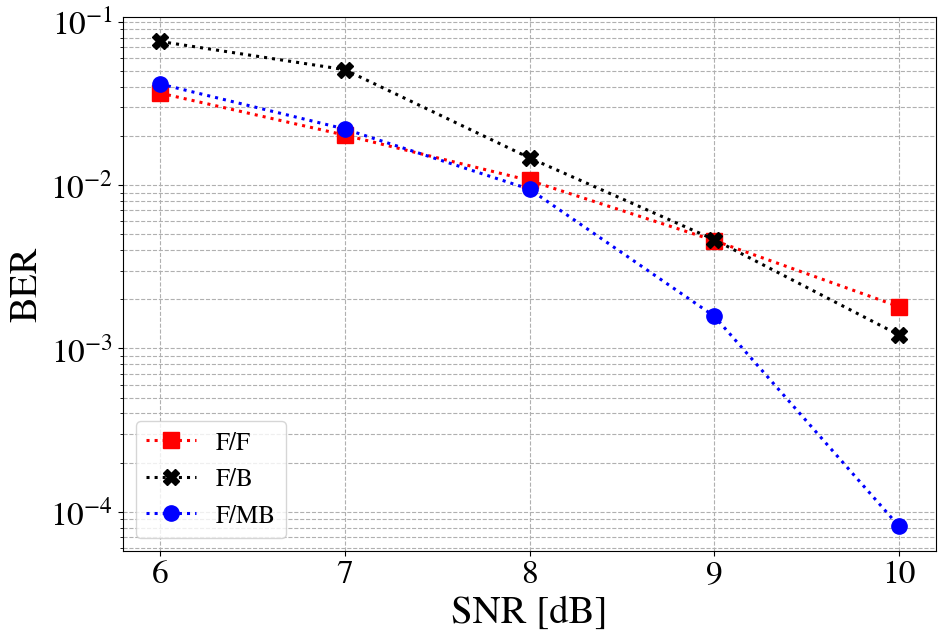}
	\caption{{Decoding performance:} \ac{ber} as a function of \ac{snr} for either a frequentist, a Bayesian or a modular Bayesian \ac{wbp} decoder. DeepSIC is trained by the frequentist method.}

\label{fig:decoding_only_ber}
\end{figure}

%We consider only frequentist detection for the next analysis. 
To evaluate modular Bayesian learning of channel decoders, we consider a \ac{bpsk} constellation.
We use $1,536$ pilot symbols, while  the rest of the hyperparameters chosen as in Subsection~\ref{subsubsec:static_linear}. For the considered setup, Fig~\ref{fig:decoding_only_ber} depicts the decoding \ac{ber} versus \ac{snr}. Large gains to decoding
are observed by employing the modular Bayesian method. This stands in contrast to the black-box Bayesian approach, which may degrade performance
at high SNR values. This behavior is in line with a similar observation regarding  DeepSIC trained in an end-to-end frequentist manner with a low number of training samples, as reported in  \cite[Fig.~10]{shlezinger2019deepSIC}.

\begin{figure*}[t]
    \centering
    \begin{subfigure}[a]{0.32\textwidth}  
    
        \centering 
        \includegraphics[width = 0.9\columnwidth]{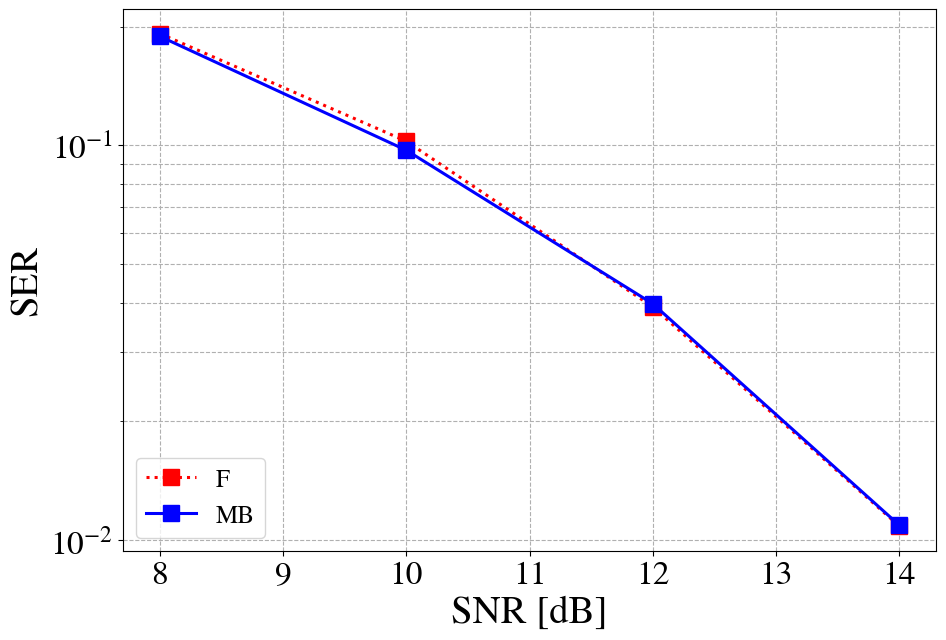}
	\caption{ \ac{ser} vs SNR}
    \label{fig:ber_and_ece_vs_snr1}
    \end{subfigure}
    \begin{subfigure}[a]{0.32\textwidth}  
    
        \centering 
        \includegraphics[width = 0.9\columnwidth]{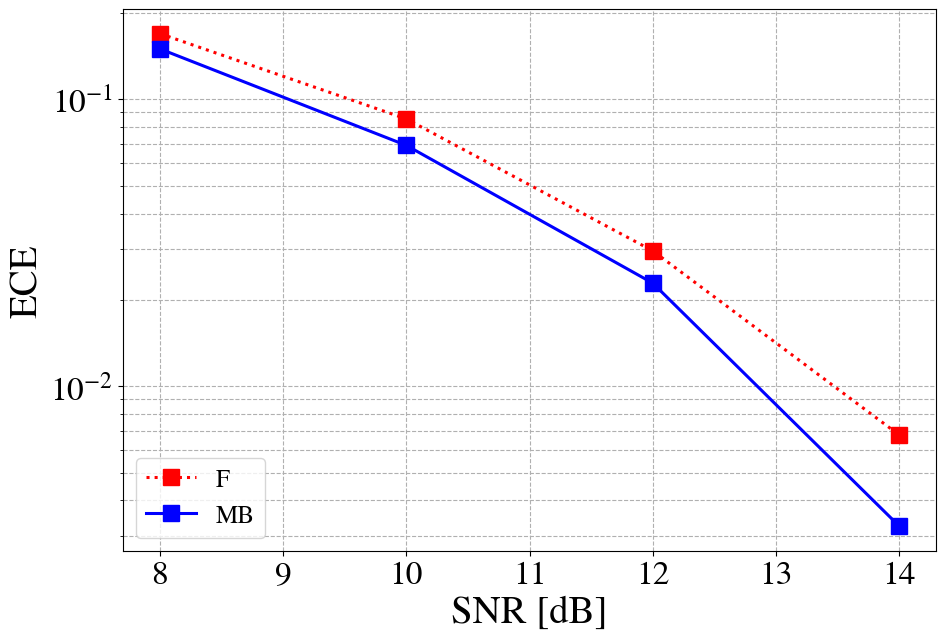}
	\caption{ECE vs SNR}
    \label{fig:ber_and_ece_vs_snr1}
    \end{subfigure}
    \hfill
    \begin{subfigure}[a]{0.32\textwidth}  
        \centering 
        \includegraphics[width = 0.9\columnwidth]{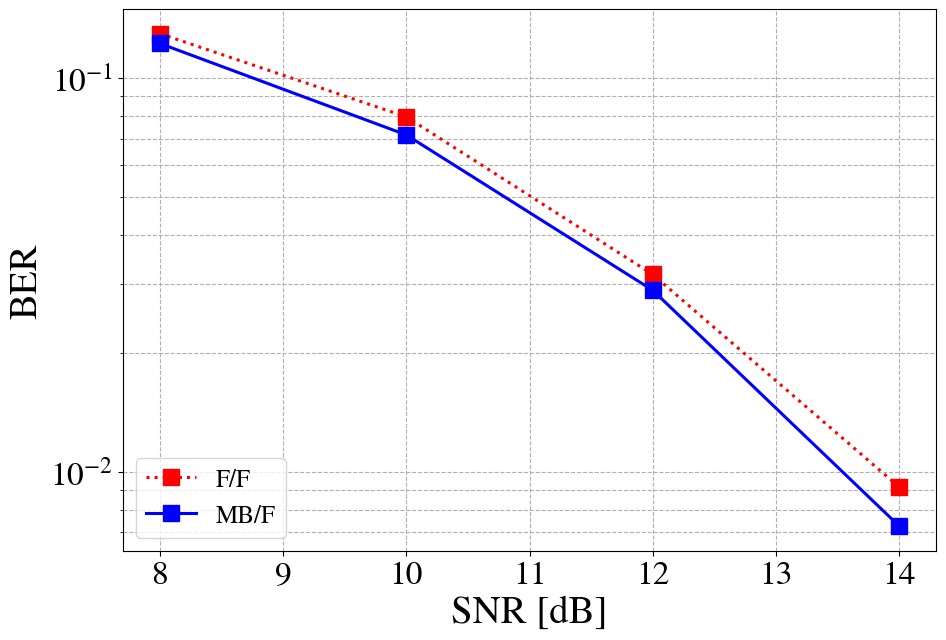}
	\caption{\ac{ber} vs SNR}
 \label{fig:ber_and_ece_vs_snr2}
    \end{subfigure}
    \caption{Analysis of decoding gain: \ac{ser}, \acs{ece}, and \ac{ber} as a function of  \ac{snr} for frequentist and modular Bayesian detector followed by a frequentist decoder.}
    \label{fig:ber_and_ece_vs_snr}
    % \vspace{-0.4cm}
\end{figure*}

 \begin{figure*}[t]
    \centering
    \begin{subfigure}[a]{0.32\textwidth}  
    
        \centering 
        \includegraphics[width = 0.9\columnwidth]{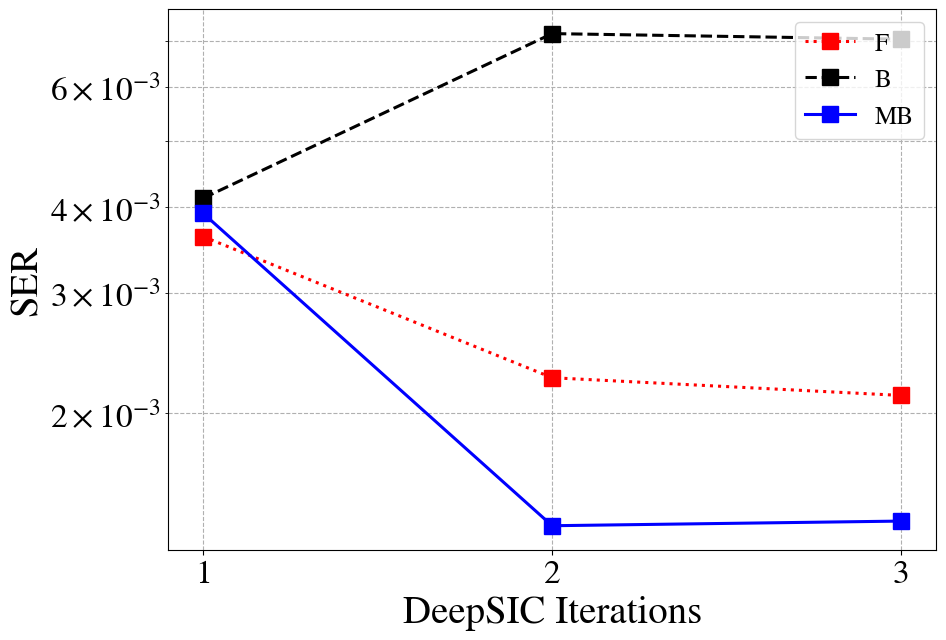}
	\caption{ \ac{ser} vs DeepSIC Iterations $(Q)$ }
    \end{subfigure}
    \begin{subfigure}[a]{0.32\textwidth}  
    
        \centering 
        \includegraphics[width = 0.9\columnwidth]{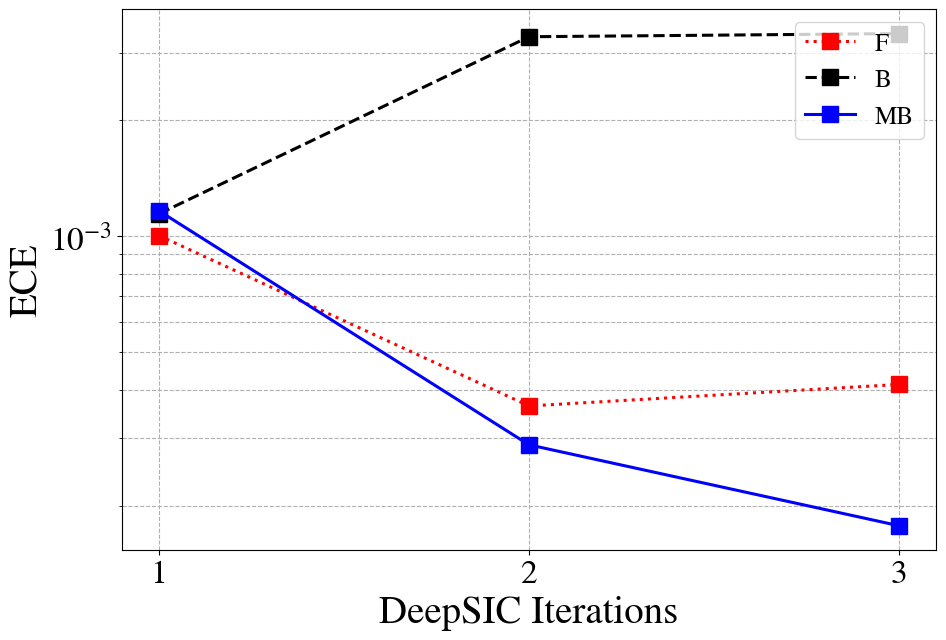}
	\caption{ECE vs  DeepSIC Iterations $(Q)$ }
    \end{subfigure}
    \hfill
    \begin{subfigure}[a]{0.32\textwidth}  
        \centering 
        \includegraphics[width = 0.9\columnwidth]{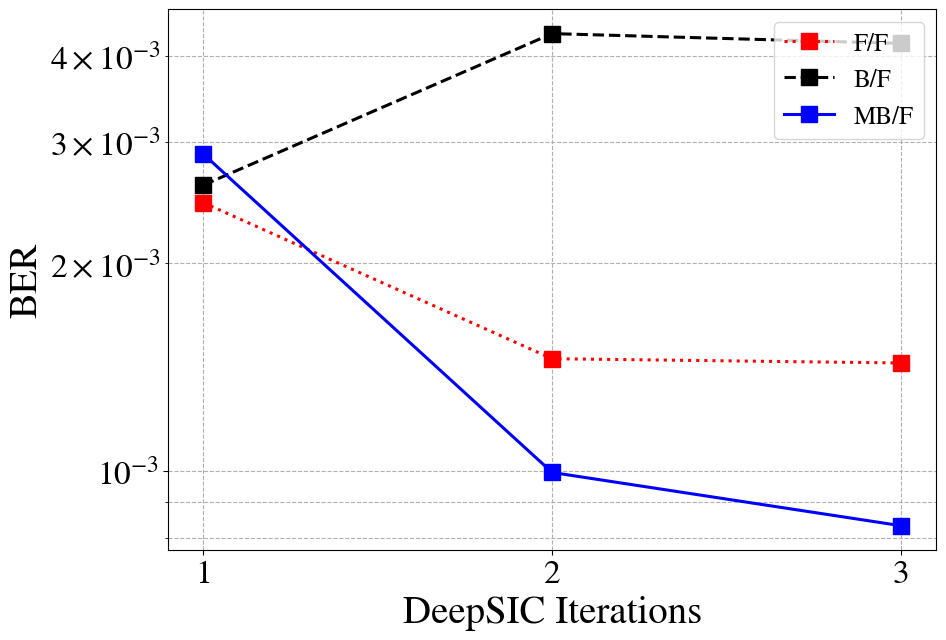}
	\caption{\ac{ber} vs DeepSIC Iterations $(Q)$ }
    \end{subfigure}
    \caption{
    Analysis of decoding gain: \ac{ser}, \acs{ece}, and \ac{ber} as a function of the number of iterations in DeepSIC for frequentist, Bayesian, and modular Bayesian detectors followed by a frequentist decoder.}
    \label{fig:effect_of_increasing_iterations}
    \vspace{-0.4cm}
\end{figure*}

%-----------------------------------
%	Gains To Downstream Tasks
%-----------------------------------
%\vspace{-0.4cm}
\subsection{Calibration and Its Effect on Decoding Gain}
\label{subsec:gains_to_downstream_tasks}

As briefly mentioned in Subsection~\ref{subsec:discussion}, applying modular Bayesian learning in equalization can also benefit decoding performance for the two following reasons: (\emph{i}) increased accuracy for hard detection, i.e., lower \ac{ser} (as reported in Subsection~\ref{subsec:detection_study}); and (\emph{ii}) increased calibration performance for soft detection which results in more reliable \acp{llr} provided to the decoders. To evaluate the individual contribution of each of these factors, we next compute the \ac{ece} \cite{naeini2015obtaining}, which is conventionally used as a scalar measure of calibration performance (see also \cite{guo2017calibration} for other calibration measures), and evaluate its contribution to soft decoding. 

% \textcolor{red}{why is ECE computed here on the symbol level while the section is about decoding? why do we first introduce ECE studies here and not for DeepSIC in the previous section? was this computed per user or the probability for equalization was evaluated over all users (multiplying the per-user distributions)?}
To formulate the \ac{ece} measure, let us partition the interval $[0,1]$ of values for the \emph{confidence level}, i.e., for the final soft output $\hat{\myVec{P}}_k$ (obtained from either frequentist learning, Bayesian learning, or modular Bayesian learning), into $R$ intervals $\{(0,\rfrac{1}{R}],...,(\rfrac{R-1}{R},1]\}$. 
% \textcolor{red}{What is $M$? how is it selected?}
Define the $r$th bin as the collection of information-symbol indices that have the corresponding confidence values lying in the $r$th interval, i.e., 
\begin{align}
    \mathcal{\Blklen}_r = \big\{ i \in \mySet{\Blklen}^{\rm info}, \forall{k} :  P(\hat{\myVec{s}}_k[i]|\myVec{y}_k[i]) \in \big(\textstyle\frac{r-1}{R}, \textstyle\frac{r}{R}\big] \big\}.
\end{align}
Accordingly, the \emph{within-bin accuracy} and \emph{within-bin confidence}  are respectively computed for each $m$th bin as
\begin{align}
    {\rm Acc}(\mathcal{\Blklen}_r) &= \frac{1}{|\mathcal{\Blklen}_r|}\sum_{i \in\mathcal{\Blklen}_r, \forall{k}}\mathbbm{1}(\hat{\myVec{s}_k}[i]=\myVec{s}_k[i]), \\
% \end{equation}
% while the within-bin confidence is computed as
% \begin{equation}
    {\rm Conf}(\mathcal{\Blklen}_r) &= \frac{1}{|\mathcal{\Blklen}_r|}\sum_{i \in \mathcal{\Blklen}_r, \forall{k}} P(\hat{\myVec{s}_k}[i]|\myVec{y}_k[i])).
\end{align}

For a perfectly calibrated model as per \eqref{eq:perfect_cal}, the  within-bin accuracy ${\rm Acc}(\mathcal{\Blklen}_r)$  must equal the within-bin confidence ${\rm Conf}(\mathcal{\Blklen}_r)$ for every bin $r=1,...,R$  given a sufficiently large set of information symbols $\mathcal{\Blklen}^\text{info}$. The \ac{ece} computes the average absolute difference between the within-bin accuracy and within-bin confidence, weighted by the number of samples that lie in each range, i.e.,
\begin{equation}
    {\rm ECE}=\sum_{r=1}^R\frac{|\mathcal{\Blklen}_r|}{\sum_{r'=1}^R |\mathcal{\Blklen}_{r'}|} \Big(|{\rm Acc}(\mathcal{\Blklen}_{r})-{\rm Conf}(\mathcal{\Blklen}_r)|\Big).
\end{equation}
The smaller the \ac{ece} is, the better calibrated the detector is. In the following analysis, we set the number of confidence bins to $R=10$

% Now that we are able to quantify calibration, we first show that in the case of only 1-iteration, the gains to the final \ac{ber} stem from the calibrated outputs of the modular Bayesian detector (``MB'') \eqref{eq:MBL_inference}, as compared to the non-calibrated outputs of the frequentist detector (``F'') \eqref{eq:freq_deepsic_soft_output}. This shows  that, as mentioned in the introduction, for any system that is designed via a cascade of multiple modules (each module not necessarily of a modular structure, e.g., a black box), calibration performance of the earlier modules can benefit the actual performance of the following modules, hence resulting in improved final performance. We will then show that whenever each module is composed of smaller modules, i.e., has modular structure rather than the black box structure, improving calibration of each smaller modules increases the calibration of the module, as well as its accuracy, hence leading to a significant performance increase as compared to  frequentist learning. Fig.~\ref{fig:ber_and_ece_vs_snr} elucidates the final \ac{ber} estimate, as well as the detector's average \ac{ser} and \ac{ece}. Note that the average \ac{ser} value per \ac{snr} is similar for both the ``F" and ``MB" detectors across all \acp{snr}. 
% Decoding is done via a frequentist-inducted

\begin{figure*}[t!]
    \centering
    \begin{subfigure}[a]{0.32\textwidth}  
    
        \centering 
        \includegraphics[width = 0.9\columnwidth]{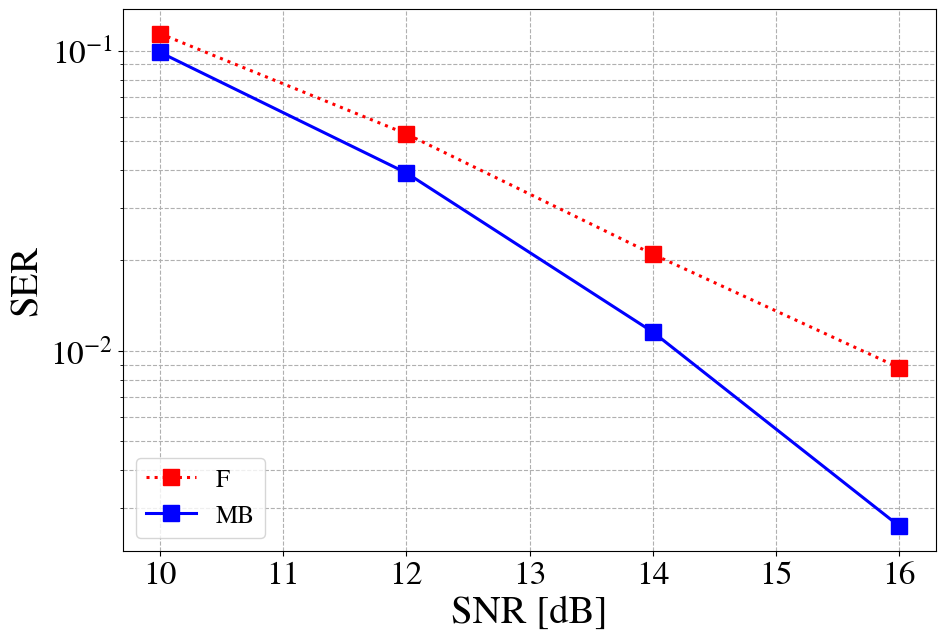}
	\caption{\ac{ser} vs SNR}
 \label{fig:synthetic_final_ser_vs_snr}
    \end{subfigure}
    \hfill
    \begin{subfigure}[a]{0.32\textwidth}  
    
        \centering 
        \includegraphics[width = 0.9\columnwidth]{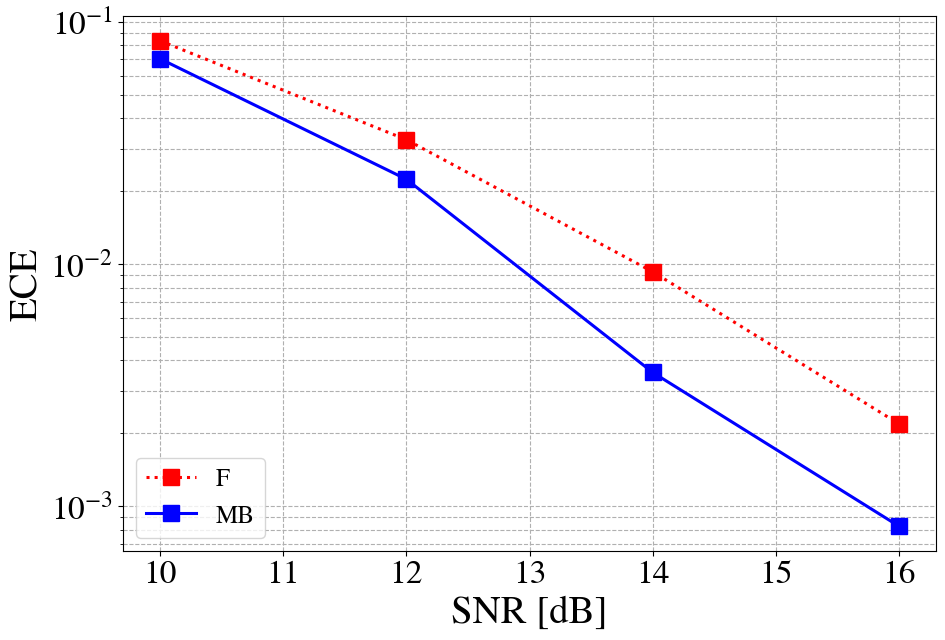}
	\caption{\ac{ece} vs SNR}
 \label{fig:synthetic_final_ece_vs_snr}
    \end{subfigure}
    \hfill
    \begin{subfigure}[a]{0.32\textwidth}  
        \centering 
        \includegraphics[width = 0.9\columnwidth]{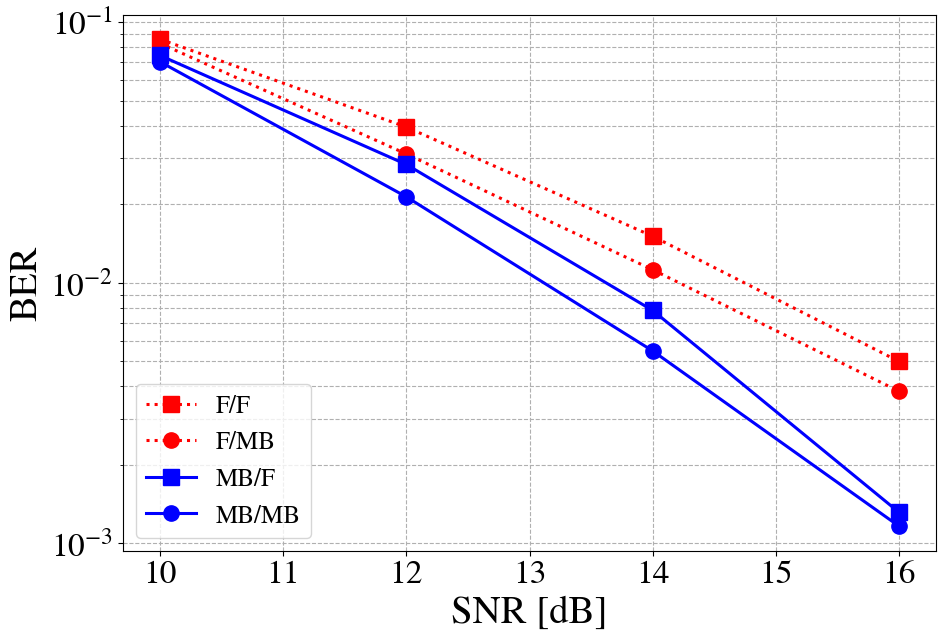}
	\caption{\ac{ber} vs SNR}
\label{fig:synthetic_final_ber_vs_snr}
    \end{subfigure}
    \caption{Overall evaluation for the synthetic channel: \ac{ser}, \ac{ece} and \ac{ber} versus \ac{snr} for the combination of either a frequentist or modular Bayesian DeepSIC followed by either a frequentist or modular Bayesian \ac{wbp} decoder.}
\label{fig:synthetic_final_comparison}

    % \vspace{-0.4cm}
\end{figure*}
 
In Fig.~\ref{fig:ber_and_ece_vs_snr}, we study the impact of the learning method for the detector by plotting the \ac{ser}, \ac{ece}, and \ac{ber} for frequentist and modular Bayesian DeepSIC. Here, we implemented DeepSIC with a single iteration $Q=1$,  followed by a standard frequentist decoder. Note that in this case, modular Bayesian DeepSIC and Bayesian DeepSIC coincide. The advantage of Bayesian learning for the detector is seen to be exclusively in terms of \ac{ece}, since the \ac{ser} is similar for frequentist and modular Bayesian DeepSIC across the considered range of \acp{snr}. The improved calibration is observed to yield a decrease in the \ac{ber} as the decoding module can benefit from the more reliable soft outputs of the Bayesian detector,  as can be clearly observed in Figs.~\ref{fig:ber_and_ece_vs_snr1}-\ref{fig:ber_and_ece_vs_snr2}. 
%This is aligned with the previous results which showed that reliable information exchange between the modules makes the modular Bayesian achieve lower \ac{ser} as compared to both frequentist and Bayesian counterpart -- now the reliable information exchange is carried out from the detector module to the decoder module.

%was similar across the range of \acp{snr}, thereby keeping the \ac{ece} as the only difference between the two methods. The ``MB'' method has outputs with lower \ac{ece}, which the \ac{wbp} decoder translated into smaller \ac{ber} values on average.

Next, we investigate how reliable information exchange between the modules within the detector 
 affects the hard-decision performance. Towards this goal, in Fig.~\ref{fig:effect_of_increasing_iterations}, we vary the modular structure of the DeepSIC detector by changing the number of iterations $Q$. We set the \ac{snr} to $15$ dB, and measure the \ac{ser}, \ac{ece}, and \ac{ber} for $Q\in\{1,2,3\}$. With a higher number of iterations $Q>1$, modular Bayesian DeepSIC is better able to exploit the modular structure, achieving both a lower \ac{ser} and a lower \ac{ece} as compared to frequentist and Bayesian DeepSIC. Its improved reliability and \ac{ser} are jointly translated at the output of the decoder into a lower \ac{ber}, as demonstrated in Fig.~\ref{fig:effect_of_increasing_iterations}.

\begin{figure*}[t!]
    \centering
    \begin{subfigure}[a]{0.32\textwidth}  
    
        \centering 
        \includegraphics[width = 0.9\columnwidth]{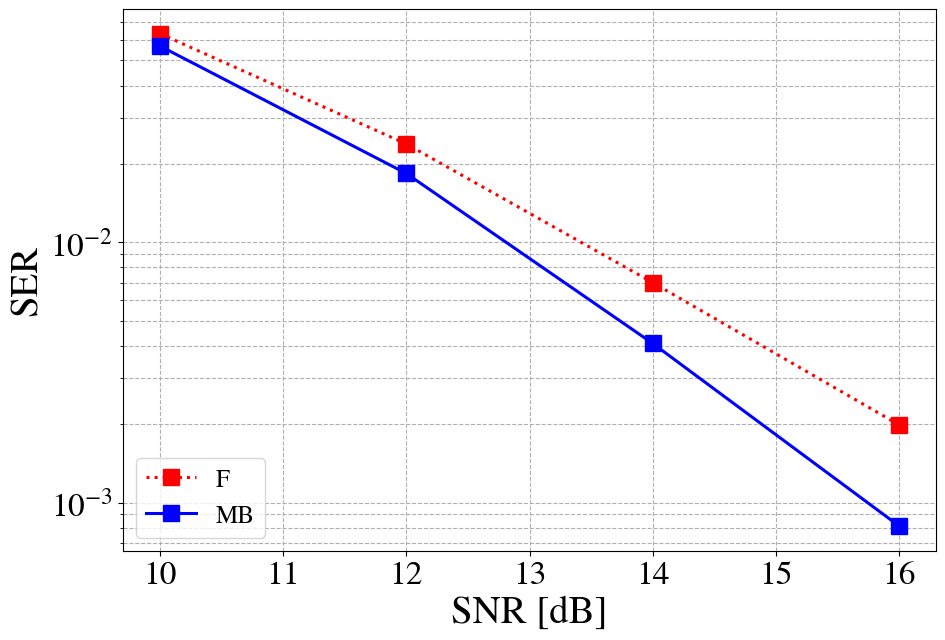}
	\caption{\ac{ser} vs SNR}
    \end{subfigure}
    \hfill
    \begin{subfigure}[a]{0.32\textwidth}  
    
        \centering 
        \includegraphics[width = 0.9\columnwidth]{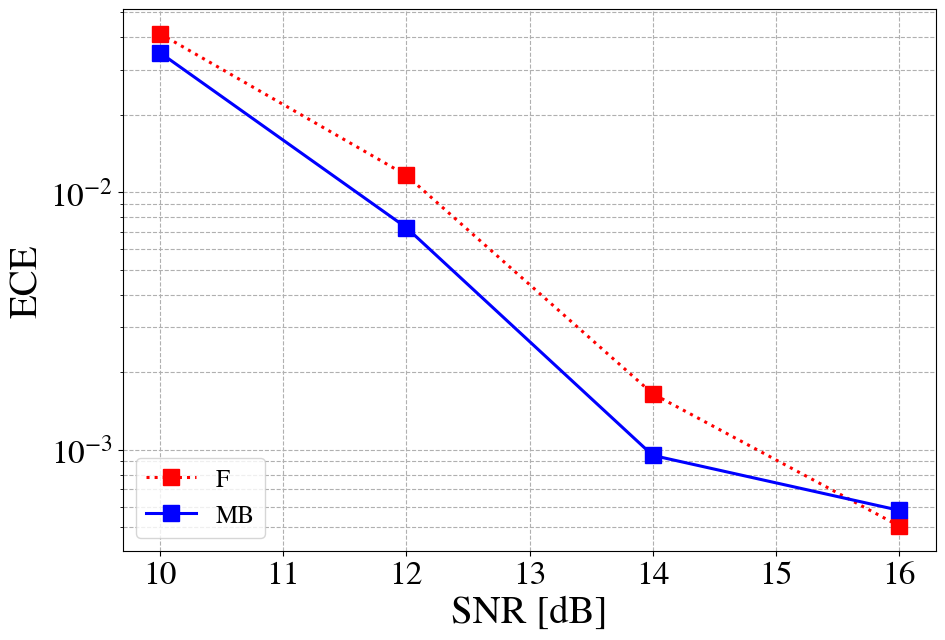}
	\caption{\ac{ece} vs SNR}
    \end{subfigure}
    \hfill
    \begin{subfigure}[a]{0.32\textwidth}  
        \centering 
        \includegraphics[width = 0.9\columnwidth]{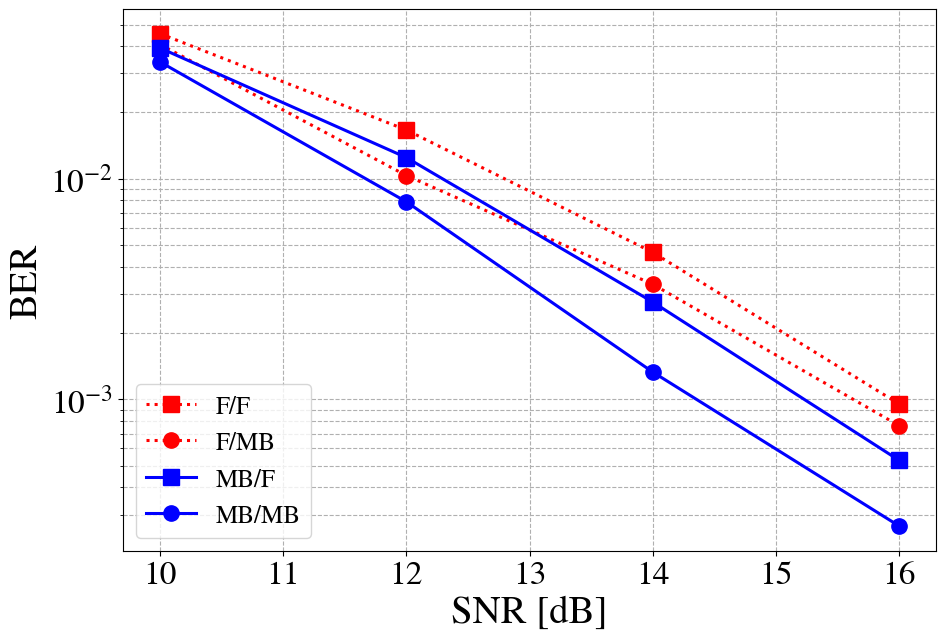}
	\caption{\ac{ber} vs SNR}
    \end{subfigure}
    \caption{Overall evaluation for the COST channel: \ac{ser}, \ac{ece} and \ac{ber} versus \ac{snr} for the combination of either a frequentist or modular Bayesian DeepSIC followed by either a frequentist or modular Bayesian \ac{wbp} decoder.}
\label{fig:cost_final_comparison}

    % \vspace{-0.4cm}
\end{figure*}
 
%-----------------------------------
%	Final Comparison
%-----------------------------------
% \vspace{-0.2cm}
\subsection{Overall Receiver Evaluation}
\label{subsec:final_comparison}
% \textcolor{red}{I am going to ask you to re-write this subsection. We need to discuss the motivation for the study, discuss each result and draw insights from it. The description below is way too minimal. }

We conclude our numerical evaluation by assessing the contribution of adopting modular Bayesian both in detection and decoding. To that end, we compare the ``F/F", ``F/MB",``MB/F" and ``MB/MB" methods for 8-PSK, with either the static synthetic channel from Subsection~\ref{subsubsec:static_linear} or the time-varying COST from Subsection~\ref{subsubsec:cost_detection} using the same hyperparameters. 

The performance achieved by the overall \ac{dnn}-aided receiver for different training schemes, when applied in the channel of Subsection~\ref{subsubsec:static_linear}, is reported in Fig.~\ref{fig:synthetic_final_comparison}. In particular, Fig.~\ref{fig:synthetic_final_ser_vs_snr} shows that the gain achievable by adopting our scheme is up to $1.5$dB in \ac{ser}. Due to this \ac{ser} gain and the improved \ac{ece} as shown in Fig.~\ref{fig:synthetic_final_ece_vs_snr}, the ``MB/F" in Fig.~\ref{fig:synthetic_final_ber_vs_snr} is shown to have similar gain in \ac{ber}. Moreover, adopting modular Bayesian in both detection and decoding (``MB/MB") is observed to have consistent gains that add up to $0.25$dB.

Fig.~\ref{fig:cost_final_comparison} reports the results for the COST channel model. We note that Fig~\ref{fig:cost_final_comparison} further reinforces the superiority of our scheme, proving the gains of the ``MB/MB" method over the ``F/F" are also apparent over a more practical channel. Under this channel, the gains of adopting modular Bayesian decoding are higher, while the gains of modular Bayesian detection are slightly smaller, with both contributing up to $0.8$dB each.
	%----------------------------------------------------------------------------------------
	%	CONCLUSIONS
	%----------------------------------------------------------------------------------------
	\section{Conclusion}
\label{sec:conclusion}
%\vspace{-0.1cm}
%\textcolor{red}{If we have no space, maybe we can omit the conclusion..}
We have introduced modular Bayesian learning, a novel integration of Bayesian learning with hybrid model-based data-driven architectures, for wireless receiver design. The proposed approach harnesses the benefits of Bayesian learning in producing well-calibrated soft outputs for each internal module, hence resulting in performance improvement via reliable information exchange across the internal modules. We have shown that applying modular Bayesian learning to the system with a cascade of modules, in which each module is composed of multiple internal modules, can yield a significant performance gain, e.g., \ac{ber} gains of up to $1.6$ dB. 

Our proposed modular Bayesian learning, integrated with model-based deep learning, is designed for deep receivers. However, its underlying design principles are expected to be useful for a broad range of tasks where modular deep architectures are employed. These include, e.g., modular \acp{dnn} designed for dynamic systems~\cite{becker2019recurrent, revach2022kalmannet} and array signal processing tasks~\cite{shmuel2023subspacenet}. 
Moreover, our deviation focuses on improving reliability via Bayesian learning. However, Bayesian learning may not improve the reliability of the modules whenever its assumption has been made wrong, e.g., prior misspecification and/or likelihood misspecification \cite{knoblauch2019generalized, masegosa2020learning, morningstar2022pacm}. This gives rise to a potential extension that replace Bayesian learning with robust Bayesian learning \cite{zecchin2023robust}, possibly combined with conformal prediction \cite{vovk2005algorithmic,angelopoulos2021gentle,cohen2023calibrating}. 
These extensions are left for future work.

%were achieved by employing the modular Bayesian modules.

%As a result, modular Bayesian learning is empirically shown to improve the overall detection accuracy, as well as the decoding capabilities, of deep receivers. For the specific channels and numerical setups, \ac{ber} gains of up to $2$ dB were achieved by employing the modular Bayesian modules.
	%----------------------------------------------------------------------------------------
	%	BIBLIOGRAPHY
	%----------------------------------------------------------------------------------------
	%\vspace{-0.2cm}
	\bibliographystyle{IEEEtran}
	\bibliography{IEEEabrv,refs}

\end{document}